\documentclass[10pt, a4paper]{article} 
\usepackage[square,numbers,sort&compress]{natbib}
\usepackage[shortlabels]{enumitem}
\usepackage{setspace}
\usepackage{subfloat}
\usepackage{mathtools}
\usepackage{amsmath,amsfonts,amssymb,amsthm,mathrsfs, bm}
\usepackage[utf8]{inputenc}
\usepackage[swedish, english]{babel}
\usepackage{csquotes}
\usepackage{subfig}
\usepackage{graphicx}
\usepackage{supertabular}                                             
\usepackage[table]{xcolor} 
\usepackage{textgreek} 
\usepackage{graphicx}
\usepackage{epstopdf}
\usepackage{microtype}
\usepackage{multirow}
\usepackage{xcolor}
\usepackage{authblk}
\usepackage{datetime2}	
\usepackage[margin=1in]{geometry} 
\usepackage{setspace}	
\theoremstyle{plain}
\newtheorem{theorem}{Theorem}[section]
\newtheorem{lemma}{Lemma}[section]

\usepackage{amsmath,amsfonts,amssymb,amsthm,mathrsfs, bm}
\usepackage{stmaryrd}

\usepackage{graphicx}
\usepackage{tikz}
\usepackage[utf8]{inputenc}
\usepackage{pgfplots} 
\usepackage{pgfgantt}
\usepackage{pdflscape}
\pgfplotsset{compat=newest} 
\pgfplotsset{plot coordinates/math parser=false}

\usepackage{endnotes}
\usepackage{commath}
\usepackage{gensymb}
\usepackage{mathtools}
\usepackage{tikz} 
\usetikzlibrary{calc,quotes,angles}
\usepackage{tkz-euclide}
\usetikzlibrary{automata,positioning} 

\usepackage{xcolor}
\usepackage{tikz,tkz-euclide,siunitx}
\usepackage{etoolbox} 

\usetikzlibrary{quotes,arrows.meta,angles,calc,shadings,positioning,babel}

\newtheorem{definition}{Definition}[section]

\newtheorem{remark}{Remark}[section]

\usepackage{endnotes}
\usepackage{commath}
\usepackage{gensymb}
\usepackage{tcolorbox}

\makeatletter
\patchcmd{\tkz@DrawLine}{\begingroup}{\begingroup\makeatletter}{}{}
\makeatother

\allowdisplaybreaks
\DeclareMathOperator{\RE}{Re}
\DeclareMathOperator{\IM}{Im}

\DeclareMathOperator{\esup}{ess\, sup}

\makeatletter
\newcommand\makebig[2]{%
  \@xp\newcommand\@xp*\csname#1\endcsname{\bBigg@{#2}}%
  \@xp\newcommand\@xp*\csname#1l\endcsname{\@xp\mathopen\csname#1\endcsname}%
  \@xp\newcommand\@xp*\csname#1r\endcsname{\@xp\mathclose\csname#1\endcsname}%
}
\makeatother

\providecommand*{\ped}[1]{%
\ensuremath{_\textnormal{#1}}}

\providecommand*{\eu}%
{\ensuremath{\mathrm{e}}}
\providecommand*{\im}%
{\ensuremath{\mathrm{i}}}
\providecommand*{\GammaF}%
{\ensuremath{\mathrm{\Gamma}}}
\providecommand*{\BetaF}%
{\ensuremath{\mathrm{\Beta}}}

\makebig{biggg} {3.0}
\makebig{Biggg} {3.5}
\makebig{bigggg}{4.0}
\makebig{Bigggg}{4.5}
\makebig{biggggg}{5.0}
\makebig{Biggggg}{5.5}
\usepackage{longtable}
\usepackage{xcolor}
\DeclareMathSymbol{\Gamma}{\mathalpha}{letters}{"00}
\DeclareMathSymbol{\Delta}{\mathalpha}{letters}{"01}
\DeclareMathSymbol{\Theta}{\mathalpha}{letters}{"02}
\DeclareMathSymbol{\Lambda}{\mathalpha}{letters}{"03}
\DeclareMathSymbol{\Xi}{\mathalpha}{letters}{"04}
\DeclareMathSymbol{\Pi}{\mathalpha}{letters}{"05}
\DeclareMathSymbol{\Sigma}{\mathalpha}{letters}{"06}
\DeclareMathSymbol{\Upsilon}{\mathalpha}{letters}{"07}
\DeclareMathSymbol{\Phi}{\mathalpha}{letters}{"08}
\DeclareMathSymbol{\Psi}{\mathalpha}{letters}{"09}
\DeclareMathSymbol{\Omega}{\mathalpha}{letters}{"0A}

\definecolor{matblue}{rgb}{0.0000,0.4470,0.7410}
\definecolor{matred}{rgb}{0.8500,0.3250,0.0980}
\definecolor{matyellow}{rgb}{0.9290,0.6940,0.1250}
\definecolor{matpurple}{rgb}{0.4940,0.1840,0.5560}
\definecolor{matgreen}{rgb}{0.4660,0.6740,0.1880}
\definecolor{matcyan}{rgb}{0.3010,0.7450,0.9330}
\definecolor{matmaroon}{rgb}{0.6350,0.0780,0.1840}

\usepackage{hyperref}
\hypersetup{
    colorlinks = true,
    linkbordercolor = {white},
            linkcolor=blue,
            bookmarks=true,
            filecolor=blue,
            urlcolor=blue,
            citecolor=blue
}

\newtcolorbox[auto counter]{modelbox}[2][]{%
  colback=white, colframe=black,
  fonttitle=\bfseries,
  title=Models~\thetcbcounter: #2,
  label=#1
}

\newcommand{\modref}[1]{Models~\ref{#1}}

\begin{document}

\title{Two-dimensional FrBD friction models for rolling contact:\\ extension to linear viscoelasticity\footnote{This document is a corrected author version of the article published in \emph{Tribology International} 220, 111953 (2026) \url{ https://doi.org/10.1016/j.triboint.2026.111953}. It incorporates corrections made after publication (see Errata) and is distributed in accordance with the CC BY licence.}}
\date{}
\author[a,b]{Luigi Romano\thanks{Corresponding author. Email: luigi.romano@liu.se.}}
\affil[a]{\footnotesize{Division of Vehicular Systems, Department of Electrical Engineering, Linköping University, SE-581 83 Linköping, Sweden}}
\affil[b]{\footnotesize{Control Systems Technology Group, Department of Mechanical Engineering, Eindhoven University of Technology, Groene Loper 1, 5612 AZ Eindhoven, the Netherlands}}

\maketitle

\begin{abstract}
This paper extends the distributed rolling contact FrBD framework to linear viscoelasticity by considering classic derivative Generalised Maxwell and Kelvin-Voigt rheological representations of the bristle element. With this modelling approach, the dynamics of the bristle, generated friction forces, and internal deformation states are described by a system of $2(n+1)$ hyperbolic \emph{partial differential equations} (PDEs), which can capture complex relaxation phenomena originating from viscoelastic behaviours. By appropriately specifying the analytical expressions for the transport and rigid relative velocity, three distributed formulations of increasing complexity are introduced, which account for different levels of spin excitation. For the linear variants, well-posedness and passivity are analysed rigorously, showing that these properties hold for any physically meaningful parametrisation. Numerical experiments complement the theoretical results by illustrating steady-state characteristics and transient relaxation effects. The findings of this paper substantially advance the FrBD paradigm by enabling a unified and systematic treatment of linear viscoelasticity.
\end{abstract}
\section*{Keywords}
Viscoelastic rolling contact; friction; friction modelling; contact mechanics; distributed parameter systems; semilinear systems

\section{Introduction}\label{intro}
Rolling contact phenomena are ubiquitous in mechanical and mechatronics engineering \cite{KinematicsMio,Flores,Flores2}, where they play a pivotal role in railway and road vehicle dynamics \cite{Knothe,KalkerBook,2000RCP,Guiggiani,Pacejka2,LibroMio,Gauterin,Gauterin2}, robotics \cite{Sphere1,Sphere2,Sphere3,Sphere4,SphereNoSlip1,SphereNoSlip2,SphereNoSlip3}, and tribology \cite{CarboneTrans,TransModel,Belt1,Frendo1,Frendo2,Frendo3,bearing1,bearing2,bearing3,bearing4,bearing5}.
Given the central role of rolling contact mechanics in engineering practice, it is unsurprising that a rich body of analytical theories has emerged to model and elucidate these phenomena. Early contributions include those of Bentall and Johnson \cite{Bentall} and Nowell and Hills \cite{Nowell}, who studied numerically the two-dimensional tractive rolling contact problem between elastically dissimilar rollers. Within the framework of linear elasticity, a comprehensive analytical treatment of rolling contact was then pioneered by Kalker, who formulated his famous theory combining rolling kinematics with a local Coulomb-Amontons friction law. Restricting attention to elastically similar cylinders undergoing longitudinal slip (or creepage) under dry friction, Kalker examined both steady and unsteady rolling behaviours in \cite{Kalker51,Kalker5,KalkerPhD}.
In the two-dimensional case, viscoelastic rolling was subsequently investigated independently by Hunter \cite{Hunter} and Goryacheva \cite{GoryachevaP}, whilst numerical approaches capable of addressing fully three-dimensional contact problems were developed by Kalker \cite{Panek1,Panek2}. Most of these studies considered viscoelastic materials characterised by a single relaxation time, which can be effectively represented by a linear elastic solid rheological model.
These pioneering contributions, comprehensively documented in standard references \cite{KalkerBook,Johnson,Goryacheva,Barber}, are now recognised as cornerstones of contact mechanics and have inspired the development of numerous numerical algorithms aimed at solving more complex rolling problems involving combined translational and spin slips \cite{Vollebregt1,Vollebregt2,Vollebregt3,Vollebregt4,Nielsen,Alfredsson}. Indeed, although elegant analytical solutions exist for steady or quasi-steady rolling \cite{GoryachevaBook}, the inherently nonlocal nature of Kalker's theory renders the derivation of closed-form results intractable, particularly when viscoelastic effects are incorporated \cite{Goryacheva4,Goryacheva1,Goryacheva2,Guler,Julia1,Julia2}.

Alternatively, Kalker's simplified theory -- also known as the \emph{brush models} in road vehicle dynamics -- offers a qualitatively valid alternative by replacing the full Cerruti-Boussinesq coupling with a Winkler-type foundation, postulating a local relationship between stresses and deformations \cite{KalkerBook,Johnson,KalkerSimp}. Following this approach, analytical solutions are reported, for instance, in \cite{Kalker4,Gross,Alonso1,Alonso2,Ciavarella1,Ciavarella2,Ciavarella3,Al-Bender,USB,LibroMio,Meccanica2,SphericalWheel,LuGreSpin,Tribology}.
Despite their success, conventional brush models exhibit a significant drawback, since they rely on an explicit partition of the contact region into adhesion and sliding zones. When rolling contact is described as a spatially distributed process, this feature severely hampers the derivation of general analytical results, complicates rigorous mathematical analysis, and limits compatibility with control and estimation frameworks. Motivated by these shortcomings, alternative formulations of rolling and spinning friction have been proposed.
For instance, the local models introduced in \cite{Zhuravlev1,Zhuravlev2,Kireenkov1,Kireenkov2} compute resultant friction forces and moments by employing rigid rolling kinematics, leading to compact expressions for global quantities. In parallel, dynamic friction models such as Dahl and LuGre \cite{Astrom1,Olsson,Astrom2} have been adapted to rolling contact problems, with an emphasis on tyre-road applications \cite{Sorine,TsiotrasConf,Tsiotras1,Tsiotras3,Deur0,Deur1,Deur2}, where viscoelastic effects may be non-negligible. These approaches do not explicitly distinguish between sticking and slipping, as they describe only the averaged deformation of bristle-like elements within the contact interface. This feature endows them with a mathematical structure that is more amenable to analysis and control design. Comparable modelling philosophies can also be pursued starting from other dynamic friction frameworks, including Leuven-type formulations \cite{Integrated,Leuven} and elastoplastic descriptions \cite{Elasto1,Elasto2}. Nevertheless, the majority of these models remain largely phenomenological and thus offer limited insight into the physical mechanisms governing friction.

More recently, lumped and distributed LuGre-inspired formulations, termed \emph{Friction with Bristle Dynamics} (FrBD), have been proposed in \cite{FrBD,FrBDroll} as a first-order approximation of a rheological model for bristle-like elements attached to sliding and rolling bodies. Owing to its physical grounding, the FrBD framework yields a coherent dynamic description of sliding phenomena under both dry and lubricated friction. In particular, the first-order version presented in \cite{FrBDroll}, renamed in this paper the rolling contact FrBD$_1$-KV model, was based on a Kelvin-Voigt (KV) rheological representation of the bristle element, which qualifies as the simplest constitutive model of a linear viscoelastic solid \cite{Rheology1,bookRheol1,bookRheol2,bookRheol3}. Consequently, the formulation presented in \cite{FrBDroll} is suitable for studying purely elastic rolling contact, or viscoelastic contact when hysteretic effects are minor. However, for highly viscoelastic materials, such as the polymers commonly used in tyres, more sophisticated rheological models are necessary \cite{JMPS1,JMPS2,JMPS3} to capture internal stress relaxation across a broad frequency spectrum \cite{Rheology2,Rheology2-3,Rheology3,Fractional1,Fractional2,Fractional3}.
Therefore, restricting itself to the classic derivative framework \cite{Rheology1,Fractional1,Fractional2,Fractional3}, the present work advances the distributed FrBD paradigm by adopting the two most general constitutive descriptions of linear viscoelasticity, namely the Generalised Maxwell (GM) and Generalised Kelvin-Voigt (GKV) elements, thereby substantially extending the modelling capability of the approach.
The governing PDEs of the new rolling contact models, collectively referred to as FrBD$_{n+1}$, are first derived following the general approach outlined in \cite{FrBDroll}. Subsequently, three distinct variants are presented and discussed in detail. For the two linear formulations, well-posedness, \emph{input-to-state stability} (ISS), and \emph{input-to-output stability} (IOS) results can be established by invoking semigroup arguments \cite{Pazy,Tanabe1,Tanabe} as in \cite{FrBDroll}, and are briefly recollected. Conversely, passivity is proved explicitly, showing that this property holds for any meaningful parametrisation of the models. The theoretical developments are complemented by numerical simulations illustrating steady-state and transient rolling contact behaviours. Force-slip relationships are computed and compared employing three different rheological descriptions, whilst transient analyses focus on relaxation dynamics. Compared to \cite{FrBDroll}, the main novelties of this work can be summarised as follows:
\begin{enumerate}[(i)]
\item Systematic extension of FrBD to arbitrary-order linear viscoelasticity via GM and GKV models,
\item Derivation of distributed hyperbolic PDEs with internal relaxation states,
\item Rigorous passivity proof for rolling contact with viscoelastic bristles.
\end{enumerate}
From an engineering perspective, the proposed FrBD$_{n+1}$ models enable the direct incorporation of experimentally identified relaxation spectra of polymers and rubbers into rolling contact simulations, without resorting to \emph{ad hoc} creep-force laws. Beyond automotive tyres, the results of this paper apply to a wide range of rolling contact elements involving viscoelastic materials, such as rubber-coated rollers used in conveying, printing, and paper-handling machinery, drive and pinch rollers in packaging and mechatronic systems, and polymer-covered calendering cylinders in plastics and rubber processing. 

The remainder of the manuscript is organised as follows. Section~\ref{sect:2Dext} presents the derivation of the FrBD$_{n+1}$ formulations, combining physically motivated assumptions with solid analytical arguments. Section~\ref{sect:models} specialises the general equations to rolling contact systems and introduces three model variants of increasing complexity. The linear formulations are then analysed in Sect.~\ref{sect:math}, with particular emphasis on well-posedness and passivity properties. Numerical results illustrating both steady and transient responses are discussed in Sect.~\ref{sect:numer}. Finally, conclusions and directions for future research are outlined in Sect.~\ref{sect:conclusion}, whilst additional details are provided in Appendices~\ref{app:param} and~\ref{app:alt}. Specifically, to facilitate practical adoption, Appendix~\ref{app:param} provides guidelines for parametrising the proposed FrBD$_{n+1}$ models from experimental frequency-domain data, whereas alternative parametrisations of certain variants are given in Appendix~\ref{app:alt}.

\subsection*{Notation}
In this paper, $\mathbb{R}$ denotes the set of real numbers; $\mathbb{R}_{>0}$ and $\mathbb{R}_{\geq 0}$ indicate the set of positive real numbers excluding and including zero, respectively. The set of positive integer numbers is indicated with $\mathbb{N}$, whereas $\mathbb{N}_{0}$ denotes the extended set of positive integers including zero, i.e., $\mathbb{N}_{0} = \mathbb{N} \cup \{0\}$.
The set of $n\times m$ matrices with values in $\mathbb{F}$ ($\mathbb{F} = \mathbb{R}$, $\mathbb{R}_{>0}$, $\mathbb{R}_{\geq0}$) is denoted by $\mathbf{M}_{n\times m}(\mathbb{F})$ (abbreviated as $\mathbf{M}_{n}(\mathbb{F})$ whenever $m=n$). $\mathbf{Sym}_n(\mathbb{R})$ represents the group of symmetric matrices with values in $\mathbb{R}$; the identity matrix on $\mathbb{R}^n$ is indicated with $\mathbf{I}_n$. A positive-definite matrix is noted as $\mathbf{M}_n(\mathbb{R}) \ni \mathbf{Q} \succ \mathbf{0}$; a positive semidefinite one as $\mathbf{M}_n(\mathbb{R}) \ni \mathbf{Q} \succeq \mathbf{0}$. 
The standard Euclidean norm on $\mathbb{R}^n$ is indicated with $\norm{\cdot}_2$; operator norms are simply denoted by $\norm{\cdot}$.
Given a domain $\Omega$ with closure $\overline{\Omega}$, $L^p(\Omega;\mathcal{Z})$ and $C^k(\overline{\Omega};\mathcal{Z})$ ($p, k \in \{1, 2, \dots, \infty\}$) denote respectively the spaces of $L^p$-integrable functions and $k$-times continuously differentiable functions on $\overline{\Omega}$ with values in $\mathcal{Z}$ (for $T = \infty$, the interval $[0,T]$ is identified with $\mathbb{R}_{\geq 0}$). In particular, $L^2(\Omega;\mathbb{R}^n)$ denotes the Hilbert space of square-integrable functions on $\Omega$ with values in $\mathbb{R}^n$, endowed with inner product $\langle \bm{\zeta}_1, \bm{\zeta}_2 \rangle_{L^2(\Omega;\mathbb{R}^n)} = \int_\Omega \bm{\zeta}_1^{\mathrm{T}}(\bm{x})\bm{\zeta}_2(\bm{x}) \dif \bm{x}$ and induced norm $\norm{\bm{\zeta}(\cdot)}_{L^2(\Omega;\mathbb{R}^n)}$. The Hilbert space $H^1(\Omega;\mathbb{R}^n)$ consists of functions $\bm{\zeta}\in L^2(\Omega;\mathbb{R}^n)$ whose weak derivative also belongs to $L^2(\Omega;\mathbb{R}^n)$. For a function $f :\Omega \to \mathbb{R}$, the sup norm is defined as $\norm{f(\cdot)}_\infty \triangleq \esup_{\Omega} \abs{f(\cdot)}$; $f : \Omega \to \mathbb{R}$ belongs to the space $L^\infty(\Omega;\mathbb{R})$ if $\norm{f(\cdot)}_\infty < \infty$. A function $f \in C^0(\mathbb{R}_{\geq 0}; \mathbb{R}_{\geq 0})$ belongs to the space $\mathcal{K}$ if it is strictly increasing and $f(0) = 0$; $f \in \mathcal{K}$ belongs to the space $\mathcal{K}_\infty$ if it is unbounded. Finally, a function $f \in C^0(\mathbb{R}_{\geq 0}^2; \mathbb{R}_{\geq 0})$ belongs to the space $\mathcal{KL}$ if $f(\cdot,t) \in \mathcal{K}$ and is strictly decreasing in its second argument, with $\lim_{t\to \infty}f(\cdot,t) = 0$. 


\section{Viscoelastic extension of the two-dimensional FrBD friction model}\label{sect:2Dext}

This section presents an extended version of the two-dimensional FrBD friction model introduced in \cite{FrBDroll}, based on more refined GM and GKV rheological descriptions of the bristle element. In particular, the proposed rheological models are described in Sect.~\ref{sect:rhelANdFr}, whilst Sect.~\ref{sect:DynamicDer} is devoted to the analytical derivation of the dynamic one.

\subsection{Viscoelastic rheological models}\label{sect:rhelANdFr}
\begin{figure}
\centering
\includegraphics[width=1\linewidth]{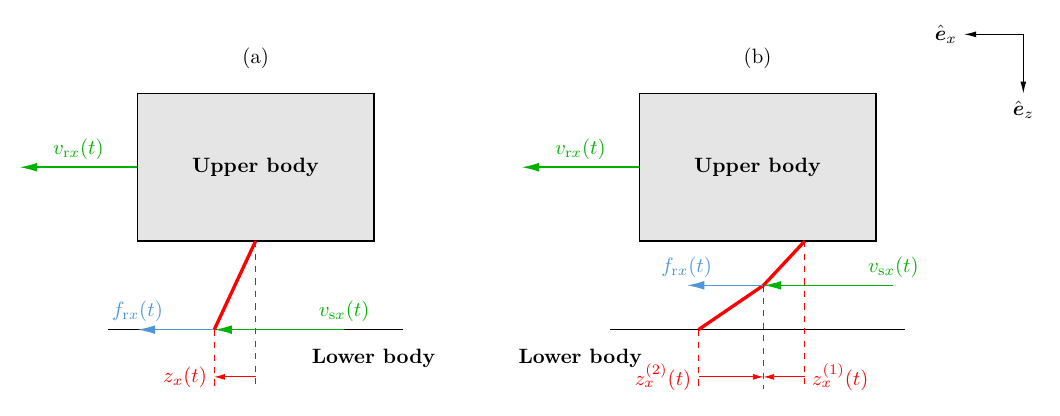} 
\caption{A schematic representation of the friction model: (a) configuration with a rigid substrate; (b) configuration with a deformable substrate. The problem is studied in a right-handed reference frame $(O;x,y,z)$ with unit vectors $(\hat{\bm{e}}_x, \hat{\bm{e}}_y, \hat{\bm{e}}_z)$.}
\label{fig:LumpModel}
\end{figure}

Following \cite{FrBDroll}, the proposed approach is elucidated by first considering the configuration illustrated in Fig.~\ref{fig:LumpModel}(a), where a viscoelastic body slides over a flat, rigid substrate. The mechanical problem is studied in a reference frame $(O;x,y,z)$ oriented as follows: the $x$-axis (longitudinal) is usually directed as the main direction of motion of the upper body, the $z$-axis (vertical) points downward, into the lower body, and the $y$-axis (lateral) is oriented so to have a right-handed system.

The rigid relative velocity of the upper body with respect to the rigid substrate (lower body) is denoted by $\mathbb{R}^2 \ni \bm{v}\ped{r} = [v_{\textnormal{r}x}\; v_{\textnormal{r}y}]^{\mathrm{T}}$. To the lower boundary of the upper body, massless bristles are attached, whose deformation is denoted by $\mathbb{R}^2 \ni \bm{z} = [z_x\; z_y]^{\mathrm{T}}$ (not to be confounded with the vertical axis $z$). The total sliding velocity between the tip of the bristle and the substrate reads
\begin{align}\label{eq:slidingS}
\bm{v}\ped{s}(\dot{\bm{z}},\bm{v}\ped{r}) = \bm{v}\ped{r}+ \dot{\bm{z}}, 
\end{align}
where Newton's notation has been adopted for the total time derivative $\dot{\bm{z}} = \od{\bm{z}}{t}$.

\begin{figure}
\centering
\includegraphics[width=0.7\linewidth]{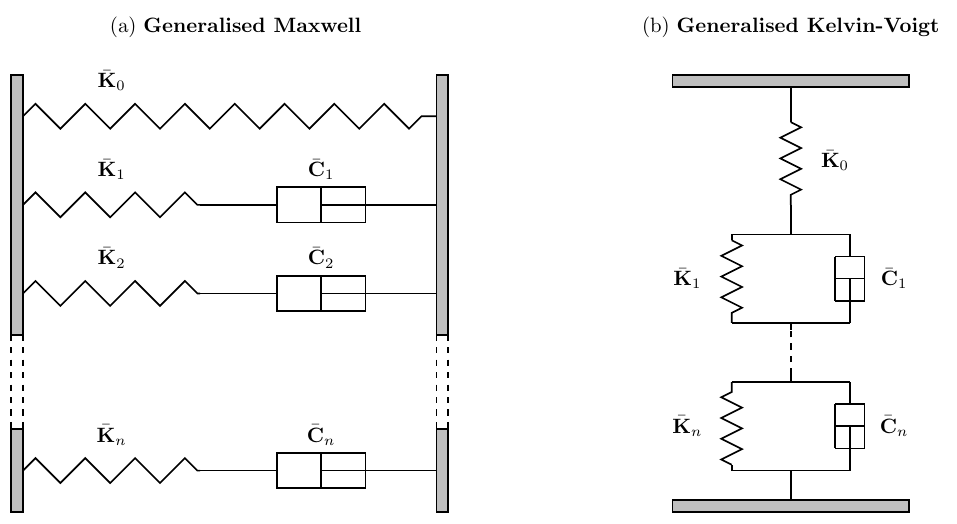} 
\caption{A schematic representation of the Generalised Maxwell (GM) and Generalised Kelvin-Voigt (GKV) rheological models. The matrix $\bar{\mathbf{K}}_0$ collects the normalised micro-stiffnesses of the zeroth element, modelled as an elastic spring. The matrices $\bar{\mathbf{K}}_i$ and $\bar{\mathbf{C}}_i$ denote the normalised micro-stiffness and micro-damping matrices of the element $i$, $i \in \{1,\dots, n\}$. For the GM, the corresponding matrix of time constants is given by $\bm{\tau}_i \triangleq \bar{\mathbf{C}}_i\bar{\mathbf{K}}_i^{-1}$, $i \in \{1,\dots,n\}$.}
\label{fig:Dashpot}
\end{figure}

By deflecting, the bristle generates a nondimensional force $\mathbb{R}^2 \ni \bm{f} = [f_x\; f_y]^{\mathrm{T}}$, which opposes the sliding motion. The (possibly dynamic) relationship between the deformation and the produced force may be specified by adopting an appropriate rheological model for the bristle element. Within the framework of linear viscoelasticity, as considered in this paper, the two most general formulations are the GM and GKV models illustrated in Fig.~\ref{fig:Dashpot}. These two models are ultimately equivalent, meaning that every GM element admits an equivalent GKV representation, and \emph{vice versa} \cite{Rheology2-3}. However, this paper considers each formulation separately to provide a unified and more comprehensive treatment of linear viscoelasticity within the theoretical FrBD framework\footnote{As mentioned in the introduction, fractional-differential models are deliberately excluded from the analysis. However, recent works have shown that fractional models can be typically reparametrised as GM or GKV ones \cite{Rheology3}, and may therefore be accommodated within the proposed framework.}.
Starting with the GM element with $n+1$ branches, depicted in Fig.~\ref{fig:Dashpot}(a), the force $\bm{f}$ may be expressed as
\begin{subequations}\label{eq:force0}
\begin{align}
\bm{f} & = \bar{\mathbf{K}}_0\bm{z} + \sum_{i = 1}^n \bm{f}_i, \label{eq:force00}\\
\dod{\bm{f}_i}{t} & = -\bm{\tau}_i^{-1}\bm{f}_i + \bar{\mathbf{K}}_i\dot{\bm{z}}, \quad i \in \{1,\dots,n\},
\end{align}
\end{subequations}
where $\mathbb{R}^2 \ni \bm{f}_i = [f_{ix}\; f_{iy}]^{\mathrm{T}}$, $i \in \{1,\dots,n\}$, denote the internal forces generated by the dissipative branches, that is, the spring-damper elements, and $\mathbf{Sym}_2(\mathbb{R}) \ni \bar{\mathbf{K}}_i \succ \mathbf{0}$, $i \in \{0,\dots,n\}$, and $\bm{\tau}_i \in \mathbf{M}_2(\mathbb{R})$, $i \in \{1,\dots,n\}$, are matrices of normalised micro-stiffness coefficients and relaxation times, respectively:
\begin{subequations}\label{eq:Sigmas1}
\begin{align}
\bar{\mathbf{K}}_i & = \begin{bmatrix} \bar{k}_{ixx} & \bar{k}_{ixy} \\ \bar{k}_{ixy} & \bar{k}_{iyy} \end{bmatrix}, \quad i \in \{0,\dots,n\},\\
\bm{\tau}_i & = \begin{bmatrix} \tau_{ixx} & \tau_{ixy} \\ \tau_{ixy} & \tau_{iyy} \end{bmatrix}, \quad i \in \{1,\dots,n\},
\end{align}
\end{subequations}
or, in the diagonal case,
\begin{subequations}\label{eq:Sigmas2}
\begin{align}
\bar{\mathbf{K}}_i & = \begin{bmatrix} \bar{k}_{ix} & 0 \\ 0 & \bar{k}_{iy} \end{bmatrix}, \quad i \in \{0,\dots,n\},\\
\bm{\tau}_i & = \begin{bmatrix} \tau_{ix} & 0 \\ 0 & \tau_{iy} \end{bmatrix}, \quad i \in \{1,\dots,n\}.
\end{align}
\end{subequations}
In particular, the relaxation matrices appearing in Eq.~\eqref{eq:force0} are given by
\begin{align}
\bm{\tau}_i \triangleq \bar{\mathbf{C}}_i\bar{\mathbf{K}}_i^{-1}, \quad i \in \{1,\dots,n\}, 
\end{align}
where $\mathbf{Sym}_2(\mathbb{R}) \ni \bar{\mathbf{C}}_i \succ \mathbf{0}$, $i \in \{1,\dots,n\}$, are matrices of normalised micro-damping coefficients:
\begin{align}\label{eq:C1}
\bar{\mathbf{C}}_i & = \begin{bmatrix} \bar{c}_{ixx} & \bar{c}_{ixy} \\ \bar{c}_{ixy} & \bar{c}_{iyy} \end{bmatrix}, \quad i \in \{1,\dots,n\},
\end{align}
or, in the diagonal case,
\begin{align}\label{eq:C2}
\bar{\mathbf{C}}_i & = \begin{bmatrix} \bar{c}_{ix} & 0 \\ 0 & \bar{c}_{iy} \end{bmatrix}, \quad i \in \{1,\dots,n\}.
\end{align}
The off-diagonal terms in $\bar{\mathbf{K}}_i$, $i \in \{0,\dots,n\}$, and $\bar{\mathbf{C}}_i$, $i \in \{1,\dots,n\}$, model coupling between longitudinal and lateral bristle deformations, which may arise from anisotropic microstructure or oriented polymer chains. In most practical applications, these matrices are diagonal, but the general formulation allows for coupled viscoelastic effects.

Equation~\eqref{eq:force0} completely characterises the internal state dynamics of a viscoelastic solid modelled by a GM element. On the other hand, by adopting a Generalised Kelvin-Voigt (GKV) representation, as illustrated in Fig.~\ref{fig:Dashpot}(b), the following relationships may be deduced: 
\begin{subequations}\label{eq:force0GKV}
\begin{align}
\bm{f} & = \bar{\mathbf{K}}_0\bm{z}-\bar{\mathbf{K}}_0\sum_{i=1}^n \bm{z}_i, \label{eq:force00GKV}\\
\bm{f} & = \bar{\mathbf{K}}_i\bm{z}_i + \bar{\mathbf{C}}_i\dot{\bm{z}}_i, \quad i \in \{1,\dots,n\},
\end{align}
\end{subequations}
where $\mathbb{R}^2 \ni \bm{z}_i = [z_{ix} \; z_{iy}]^{\mathrm{T}}$, $i \in \{1,\dots,n\}$, indicate the internal deformations of the dissipative branches, and $\mathbf{Sym}_2(\mathbb{R}) \ni \bar{\mathbf{K}}_i \succ \mathbf{0}$, $i \in \{0,\dots,n\}$ and $\mathbf{Sym}_2(\mathbb{R}) \ni \bar{\mathbf{C}}_i \succ \mathbf{0}$, $i \in \{1,\dots,n\}$, have the same meaning as before.
Differentiating and manipulating Eqs.~\eqref{eq:force0} and~\eqref{eq:force0GKV}, a general constitutive equation may be derived in differential form as\footnote{Equation~\eqref{eq:rheol1_diff} provides an equivalence between the GM and GKV representations. In this context, it is worth observing that the matrices $\bar{\mathbf{K}}_i$, $i \in \{0,\dots,n\}$, and $\bar{\mathbf{C}}_i, \bm{\tau}_i$, $i \in \{1,\dots,n\}$, appearing in Eqs.~\eqref{eq:force0} and~\eqref{eq:force0GKV} are supposed to have constant coefficients in space (see also Appendix~\ref{app:param}). This is a standard assumption in the literature \cite{Sorine,TsiotrasConf,Tsiotras1,Tsiotras3,Deur0,Deur1,Deur2}.}
\begin{align}\label{eq:rheol1_diff}
\bm{f}+\sum_{i=1}^n\mathbf{\Gamma}_i\dod[i]{\bm{f}}{t} = \sum_{i=0}^n \mathbf{\Sigma}_i\dod[i]{\bm{z}}{t},
\end{align}
where the matrices $\mathbf{\Gamma}_i\in \mathbf{M}_2(\mathbb{R})$, $i \in \{1,\dots,n\}$, and $\mathbf{\Sigma}_i \in \mathbf{M}_2(\mathbb{R})$, $i \in \{0,\dots,n\}$, may be inferred from the $\bar{\mathbf{K}}_i$'s and $\bm{\tau}_i$'s appearing in Eq.~\eqref{eq:force0} when considering the GM representation, and from the $\bar{\mathbf{K}}_i$'s and $\mathbf{C}_i$'s in~\eqref{eq:force0GKV} when using the GKV model. Their specification is not essential for what follows, and therefore an explicit parametrisation of the $\mathbf{\Gamma}_i$'s and $\mathbf{\Sigma}_i$'s in terms of the models' physical coefficients is not provided. Besides, it should be mentioned that inferring interconversions of a general nature for every $n \in \mathbb{N}$ is a rather arduous task \cite{Rheology2,Rheology2-3}. For $n=1$ and $n=2$, closed-form expressions are reported, for instance, in \cite{Rheology2-3}, but omitted here.
 
What is instead important to clarify in the context of the present paper is that equations formally identical to~\eqref{eq:force0} and~\eqref{eq:rheol1} may also be deduced when considering the scenario depicted in Fig.~\ref{fig:LumpModel}(b), where the upper and lower bodies exhibit a similar viscoelastic behaviour (see \cite{FrBDroll} for a comprehensive justification of this claim). 

Equations~\eqref{eq:force0} and ~\eqref{eq:rheol1} provide the main physical ingredients required to derive the extended FrBD model, as explained in the following Sect.~\ref{sect:DynamicDer}.

\subsection{Dynamic friction models derivation}\label{sect:DynamicDer}
Combining the viscoelastic rheological models detailed in Sect.~\ref{sect:rhelANdFr} with a static law for the friction coefficient, the equations governing the bristle dynamics may be inferred via an application of Theorem~\ref{thm:Theorem1} below.

\begin{theorem}[Edwards \cite{Edwards}]\label{thm:Theorem1}
Suppose that the mapping $\bm{H} : \mathbb{R}^{m+n}\to \mathbb{R}^n$ is $C^1$ in a neighbourhood of a point $(\bm{x}^\star,\bm{y}^\star)$, where $\bm{H}(\bm{x}^\star,\bm{y}^\star) = \bm{0}$. If the Jacobian matrix $\nabla_{\bm{y}}\bm{H}(\bm{x}^\star,\bm{y}^\star)^{\mathrm{T}}$ is nonsingular, there exist a neighbourhood $\mathcal{X}$ of $\bm{x}^\star$ in $\mathbb{R}^m$, a neighbourhood $\mathcal{Y}$ of $(\bm{x}^\star,\bm{y}^\star)$ in $\mathbb{R}^{m+n}$, and a mapping $\bm{h} \in C^1(\mathcal{X};\mathbb{R}^n)$ such that $\bm{y} = \bm{h}(\bm{x})$ solves the equation $\bm{H}(\bm{y},\bm{x}) = \bm{0}$ in $\mathcal{Y}$. 
In particular, the implicitly defined mapping $\bm{h}(\cdot)$ is the limit of the sequence $\{\bm{h}_k\}_{ k\in \mathbb{N}_0}^\infty$ of the successive approximations inductively defined by
\begin{subequations}
\begin{align}
\bm{h}_{k+1}(\bm{x}) & = \bm{h}_k(\bm{x}) - \nabla_{\bm{y}}\bm{H}(\bm{x}^\star,\bm{y}^\star)^{-\mathrm{T}}\bm{H}\bigl(\bm{x},\bm{h}_k(\bm{x})\bigr), \\
 \bm{h}_0(\bm{x}) & = \bm{y}^\star,
\end{align}
\end{subequations}
for $\bm{x} \in \mathcal{X}$.
\end{theorem}
By invoking Theorem~\ref{thm:Theorem1}, it is possible to proceed with the derivation of the new dynamic friction models following a similar rationale as in \cite{FrBDroll}. The main steps are repeated here for self-containment. First, Eq.~\eqref{eq:rheol1_diff} is reinterpreted algebraically, that is,
\begin{align}\label{eq:rheol1}
\bm{f}\Biggl(\dod[n]{\bm{z}}{t},\dots,\dod{\bm{z}}{t},\bm{z}, \dod[n]{\bm{f}}{t},\dots,\dod{\bm{f}}{t}\Biggr) = \sum_{i=0}^n \mathbf{\Sigma}_i\dod[i]{\bm{z}}{t}-\sum_{i=1}^n\mathbf{\Gamma}_i\dod[i]{\bm{f}}{t}.
\end{align}
As a second step, it should be observed that, in the absence of inertial effects, the normalised bristle force per unit of length must counteract the friction force per unit of normal load $p \in \mathbb{R}_{>0}$ acting on its tip, $\mathbb{R}^2 \ni \bm{f}\ped{r} = [f_{\textnormal{r}x}\; f_{\textnormal{r}y}]^{\mathrm{T}}$. Inspired by \cite{FrBDroll}, the latter may be modelled as
 \begin{align}\label{eq:frModified}
\bm{f}\ped{r}(\bm{v}\ped{s}) =- \dfrac{\mathbf{M}^2(\bm{v}\ped{s})\bm{v}\ped{s}}{\norm{\mathbf{M}(\bm{v}\ped{s})\bm{v}\ped{s}}_{2,\varepsilon}},
\end{align}
where
\begin{align}\label{eq;matrixM}
\mathbf{M}(\bm{v}\ped{s}) = \begin{bmatrix} \mu_{xx}(\bm{v}\ped{s}) &  \mu_{xy}(\bm{v}\ped{s}) \\ \mu_{xy}(\bm{v}\ped{s})  & \mu_{yy}(\bm{v}\ped{s})\end{bmatrix}
\end{align}
is a symmetric, positive definite matrix of friction coefficients, i.e., $\mathbf{Sym}_2(\mathbb{R}) \ni \mathbf{M}(\bm{y}) \succ \mathbf{0}$ for all $\bm{y} \in \mathbb{R}^2$, and $\varepsilon \in \mathbb{R}_{\geq 0}$ represents a regularisation parameter, and $\norm{\cdot}_{2,\varepsilon} \in C^0(\mathbb{R}^2;\mathbb{R}_{\geq 0})$ is a regularisation of the Euclidean norm $\norm{\cdot}_2$ for $\varepsilon\in \mathbb{R}_{>0}$, often converging uniformly to $\norm{\cdot}_2$ in $C^0(\mathbb{R}^2;\mathbb{R}_{\geq 0})$ for $\varepsilon \to 0$ (e.g., $\norm{\bm{y}}_{2,\varepsilon }= \sqrt{\norm{\bm{y}}_2^2 +\varepsilon}$), and with $\norm{\cdot}_{2,\varepsilon} \in C^1(\mathbb{R}^2;\mathbb{R}_{\geq 0})$ for $\varepsilon \in \mathbb{R}_{>0}$.

Accordingly, starting with Eqs.~\eqref{eq:rheol1} and~\eqref{eq:frModified}, and adopting the notation of Theorem~\ref{thm:Theorem1}, equating Eqs.~\eqref{eq:rheol1} and~\eqref{eq:frModified} gives
\begin{align}\label{eq:H}
\begin{split}
& \bm{H}\Biggl(\dod[n]{\bm{z}}{t},\dots,\dod{\bm{z}}{t},\bm{z}, \dod[n]{\bm{f}}{t},\dots,\dod{\bm{f}}{t},\bm{v}\ped{r}\Biggr) \\
& \quad = \bm{f}\Biggl(\dod[n]{\bm{z}}{t},\dots,\dod{\bm{z}}{t},\bm{z}, \dod[n]{\bm{f}}{t},\dots,\dod{\bm{f}}{t}\Biggr) - \bm{f}\ped{r}\bigl(\bm{v}\ped{s}(\dot{\bm{z}},\bm{v}\ped{r})\bigr) = \bm{0}, \quad t \in (0,T).
\end{split}
\end{align}
In the sliding regime, where $\norm{\dot{\bm{z}}}_2 \ll \norm{\bm{v}\ped{r}}_2$, Eq.~\eqref{eq:H} may be approximated by invoking Theorem~\ref{thm:Theorem1} with $\bm{x} = (\bm{z},\od[2]{\bm{z}}{t},\dots,\od[n]{\bm{z}}{t},\od{\bm{f}}{t},\dots,\od[n]{\bm{f}}{t},\bm{v}\ped{r})$ and $\bm{y} = \dot{\bm{z}} = \od{\bm{z}}{t}$, yielding
\begin{align}\label{eq:zk01}
\begin{split}
\dod{\bm{z}_{k+1}}{t} & = \dod{\bm{z}_k}{t}-\nabla_{\bm{\dot{z}}}\bm{H}\Biggl(\dod[n]{\bm{z}}{t}^\star,\dots,\dod{\bm{z}}{t}^\star,\bm{z}^\star, \dod[n]{\bm{f}}{t}^\star,\dots,\dod{\bm{f}}{t}^\star,\bm{v}\ped{r}^\star\Biggr)^{-\mathrm{T}}\\
& \quad \times \bm{H}\Biggl(\dod[n]{\bm{z}_k}{t},\dots,\dod{\bm{z}}{t},\bm{z}, \dod[n]{\bm{f}}{t},\dots,\dod{\bm{f}}{t},\bm{v}\ped{r}\Biggr), \quad k \in \mathbb{N}_0.
\end{split}
\end{align}
In turn, committing the additional approximation \cite{FrBD,FrBDroll,Rill}
\begin{align}\label{eq:nablaH}
\begin{split}
& \nabla_{\bm{\dot{z}}}\bm{H}\Biggl(\dod[n]{\bm{z}}{t},\dots,\dod{\bm{z}}{t},\bm{z}, \dod[n]{\bm{f}}{t},\dots,\dod{\bm{f}}{t},\bm{v}\ped{r}\Biggr)^{\mathrm{T}}   \approx \mathbf{\Sigma}_1 + \dfrac{\mathbf{M}^2\bigl(\bm{v}\ped{s}(\dot{\bm{z}},\bm{v}\ped{r})\bigr)}{\norm{\mathbf{M}\bigl(\bm{v}\ped{s}(\dot{\bm{z}},\bm{v}\ped{r})\bigr)\bm{v}\ped{s}(\dot{\bm{z}},\bm{v}\ped{r})}_{2,\varepsilon}},
\end{split}
\end{align}
recalling Eqs.~\eqref{eq:force0} and~\eqref{eq:rheol1}, and truncating Eq.~\eqref{eq:zk01} at $k=1$ provides, for an initial guess $\dot{\bm{z}}_0 = \bm{0}$,
\begin{align}\label{eq:ODEz}
\dot{\bm{z}}(t) = -\mathbf{M}^{-2}\bigl(\bm{v}\ped{r}(t)\bigr)\norm{\mathbf{M}\bigl(\bm{v}\ped{r}(t)\bigr)\bm{v}\ped{r}(t)}_{2,\varepsilon}\bm{f}(t)-\bm{v}\ped{r}(t), \quad t \in (0,T).
\end{align}
The above ODE~\eqref{eq:ODEz} is valid regardless of the assumed rheological representation of the bristle element. However, the expression for the total normalised force $\bm{f}(t)$ appearing in Eq.~\eqref{eq:ODEz} depends on the specific set of constitutive relationships, namely~\eqref{eq:force0} or~\eqref{eq:force0GKV}. In particular, the GM element described by Eq.~\eqref{eq:force0} yields the following ODE system: 
\begin{subequations}\label{eq:ODEModel}
\begin{align}
& \dot{\bm{z}}(t) = -\mathbf{M}^{-2}\bigl(\bm{v}\ped{r}(t)\bigr)\norm{\mathbf{M}\bigl(\bm{v}\ped{r}(t)\bigr)\bm{v}\ped{r}(t)}_{2,\varepsilon}\bigggl( \bar{\mathbf{K}}_0\bm{z}(t) + \sum_{i=1}^n \bm{f}_i(t)\bigggr)-\bm{v}\ped{r}(t), \\
& \dot{\bm{f}}_i(t) = -\bm{\tau}_i^{-1}\bm{f}_i(t) + \bar{\mathbf{K}}_i\dot{\bm{z}}(t), \quad t \in (0,T), \; i \in \{1,\dots,n\},\\
& \bm{z}(0) = \bm{z}_0, \; \bm{f}_i(0) = \bm{f}_{i,0}, \quad i \in \{1,\dots,n\},
\end{align}
\end{subequations}
whereas employing the GKV~\eqref{eq:force0GKV} provides
\begin{subequations}\label{eq:ODEModelKV}
\begin{align}
& \dot{\bm{z}}(t) = -\mathbf{M}^{-2}\bigl(\bm{v}\ped{r}(t)\bigr)\norm{\mathbf{M}\bigl(\bm{v}\ped{r}(t)\bigr)\bm{v}\ped{r}(t)}_{2,\varepsilon}\bar{\mathbf{K}}_0\bigggl(\bm{z}(t) - \sum_{i=1}^n \bm{z}_i(t)\bigggr)-\bm{v}\ped{r}(t), \\
& \dot{\bm{z}}_i(t) = -\bar{\mathbf{C}}_i^{-1}\bar{\mathbf{K}}_i\bm{z}_i(t) + \bar{\mathbf{C}}_i^{-1}\bar{\mathbf{K}}_0\bigggl(\bm{z}(t) - \sum_{i=1}^n \bm{z}_i(t)\bigggr), \quad t \in (0,T), \; i \in \{1,\dots,n\},\\
& \bm{z}(0) = \bm{z}_0, \; \bm{z}_i(0) = \bm{z}_{i,0}, \quad i \in \{1,\dots,n\},
\end{align}
\end{subequations}
The ODE system~\eqref{eq:ODEModel} describes the dynamics of the bristle $\bm{z}(t) \in \mathbb{R}^2$ and element forces $\bm{f}_i(t) \in \mathbb{R}^2$, $i \in \{1,\dots,n\}$, depending on the rigid relative velocity input $\bm{v}\ped{r}(t)\in \mathbb{R}^2$. In this paper, it is referred to as the FrBD$_{n+1}$-GM model, where $n+1$ indicates the dynamical order\footnote{More precisely, since all the state variables are in $\mathbb{R}^2$, the actual order is $2(n+1)$. However, here $n+1$ indicates the order in one dimension, which also coincides with the number of branches employed in the GM model.}, and the suffix GM clarifies that it corresponds to a Generalised Maxwell rheological element. In the ODE~\eqref{eq:ODEModelKV}, the dynamics of the bristle is instead coupled to that of the deformations $\bm{z}_i(t) \in \mathbb{R}^2$, $i \in \{1,\dots,n\}$. The model described by Eq.~\eqref{eq:ODEModelKV} is named in this paper the FrBD$_{n+1}$-GKV, to clarify that it is obtained by adopting a GKV rheological description of the bristle element. Together, the FrBD$_{n+1}$-GM and FrBD$_{n+1}$-GKV are collectively called FrBD$_{n+1}$ models. It is worth emphasising that the FrBD$_{n+1}$-GM and FrBD$_{n+1}$-GKV are dynamically equivalent, since there always exists an interconversion between the GM and GKV descriptions, as implied by the general constitutive relationship~\eqref{eq:rheol1_diff}. Essentially, they correspond to two alternative realisations of the viscoelastic friction dynamics, albeit the specific interpretation of their internal state variables may differ in practice. In this context, it is interesting to draw some considerations about the stationary solutions of Eqs.~\eqref{eq:ODEModel} and~\eqref{eq:ODEModelKV}. Starting with the steady-state bristle force, both formulations give
\begin{align}\label{eq:fStat}
\bm{f}(\bm{v}\ped{r}) = \bm{f}\ped{r}(\bm{v}\ped{r})  = -\dfrac{\mathbf{M}^2(\bm{v}\ped{r}\bigr)\bm{v}\ped{r}}{\norm{\mathbf{M}(\bm{v}\ped{r})\bm{v}\ped{r}}_{2,\varepsilon}},
\end{align}
which agrees with the expression deduced for the FrBD$_{1}$-KV models developed in \cite{FrBDroll}. It is worth noting that, in contrast to LuGre, a minus sign appears in Eq.~\eqref{eq:fStat}, which reflects the nature of the sliding problem, and is consistent with the formulae reported in \cite{FrBDroll,Antali}.

Concerning instead the internal variables, the FrBD$_{n+1}$-GM model~\eqref{eq:ODEModel} yields
\begin{align}
\bm{z}(\bm{v}\ped{r}) = -\bar{\mathbf{K}}_0^{-1}\dfrac{\mathbf{M}^2(\bm{v}\ped{r}\bigr)\bm{v}\ped{r}}{\norm{\mathbf{M}(\bm{v}\ped{r})\bm{v}\ped{r}}_{2,\varepsilon}}, \quad \bm{f}_i(\bm{v}\ped{r}) = \bm{0}, \quad i \in \{1,\dots,n\},
\end{align}
which coincides with the stationary solution of the FrBD$_1$-KV model, whereas the FrBD$_{n+1}$-GKV variant~\eqref{eq:ODEModelKV} provides
\begin{align}
\bm{z}(\bm{v}\ped{r}) = -\sum_{i=0}^n\bar{\mathbf{K}}_i^{-1}\dfrac{\mathbf{M}^2(\bm{v}\ped{r}\bigr)\bm{v}\ped{r}}{\norm{\mathbf{M}(\bm{v}\ped{r})\bm{v}\ped{r}}_{2,\varepsilon}}, \quad \bm{z}_i(\bm{v}\ped{r}) = -\bar{\mathbf{K}}_i^{-1}\dfrac{\mathbf{M}^2(\bm{v}\ped{r}\bigr)\bm{v}\ped{r}}{\norm{\mathbf{M}(\bm{v}\ped{r})\bm{v}\ped{r}}_{2,\varepsilon}}, \quad i \in \{1,\dots,n\},
\end{align}
which is consistent with the parallel structure of the GKV element.

For $n = 0$ and $n =1$ and suitable parametrisations, respectively, the FrBD$_{n+1}$ models~\eqref{eq:ODEModel} and~\eqref{eq:ODEModelKV} reduce to the Dahl and FrBD$_2$-SLS formulations, which are based on a single elastic element, and on a Standard Linear Solid (SLS) or Zener model. In turn, it may be easily concluded that the FrBD$_2$-SLS description further simplifies into the FrBD$_1$-KV model introduced in \cite{FrBDroll} for a specific parametrisation of the matrices $\bar{\mathbf{K}}_1$ and $\bm{\tau}_1$, generalising all the FrBD formulations currently available in the literature (see Appendix~\ref{app:alt}).

A second interesting consideration pertains to the dissipative nature of the model, which should be evaluated considering the true sliding velocity $\bm{v}\ped{s}(\dot{\bm{z}},\bm{v}\ped{r})$. In particular, Eq.~\eqref{eq:ODEz} yields
\begin{align}
\bm{v}\ped{s}\bigl(\dot{\bm{z}}(t),\bm{v}\ped{r}(t)\bigr) = \bm{v}\ped{r}(t) + \dot{\bm{z}}(t) = -\mathbf{M}^{-2}\bigl(\bm{v}\ped{r}(t)\bigr)\norm{\mathbf{M}\bigl(\bm{v}\ped{r}(t)\bigr)\bm{v}\ped{r}(t)}_{2,\varepsilon}\bm{f}(t), \quad t \in [0,T].
\end{align}
Consequently, for $p \in C^0([0,T]; \mathbb{R}_{\geq 0})$, the dissipated power reads
\begin{align}
-p(t)\bm{f}^{\mathrm{T}}(t)\bm{v}\ped{s}\bigl(\dot{\bm{z}}(t),\bm{v}\ped{r}(t)\bigr) =p(t)\bm{f}^{\mathrm{T}}(t)\mathbf{M}^{-2}\bigl(\bm{v}\ped{r}(t)\bigr)\norm{\mathbf{M}\bigl(\bm{v}\ped{r}(t)\bigr)\bm{v}\ped{r}(t)}_{2,\varepsilon}\bm{f}(t) \geq 0, \quad t \in [0,T].
\end{align}
Equations~\eqref{eq:ODEModel} and~\eqref{eq:ODEModelKV} appear therefore to preserve the natural properties of the friction force~\eqref{eq:frModified}. They are, however, still finite-dimensional, and cannot accurately capture spatially-varying rolling contact phenomena that evolve over a finite area. The distributed-parameter extension of Eqs.~\eqref{eq:ODEModel} and~\eqref{eq:ODEModelKV} is worked out next in Sect.~\ref{sect:models}.

\section{Distributed rolling contact models}\label{sect:models}
This section introduces three rolling contact models of varying complexity that can be derived from Eqs.~\eqref{eq:ODEModel} and~\eqref{eq:ODEModelKV}. Specifically, the next Sect.~\ref{sect:gener} discusses some generalities, whereas the three formulations are detailed in Sect.~\ref{sect:modelRollSimple}.


\subsection{Generalities}\label{sect:gener}
Following the discussion initiated in \cite{FrBDroll}, three important aspects that are first clarified concern the formulation of Eqs.~\eqref{eq:ODEModel} and~\eqref{eq:ODEModelKV} as PDE systems in the Eulerian setting, the prescription of appropriate \emph{boundary conditions} (BCs), and the calculation of the tangential forces and vertical moment. These are addressed respectively in Sects.~\ref{sect:eulerian},~\ref{sect:BC}, and~\ref{ect:forces}.

\subsubsection{Eulerian approach and transport velocity}\label{sect:eulerian}
\begin{figure}
\centering
\includegraphics[width=1\linewidth]{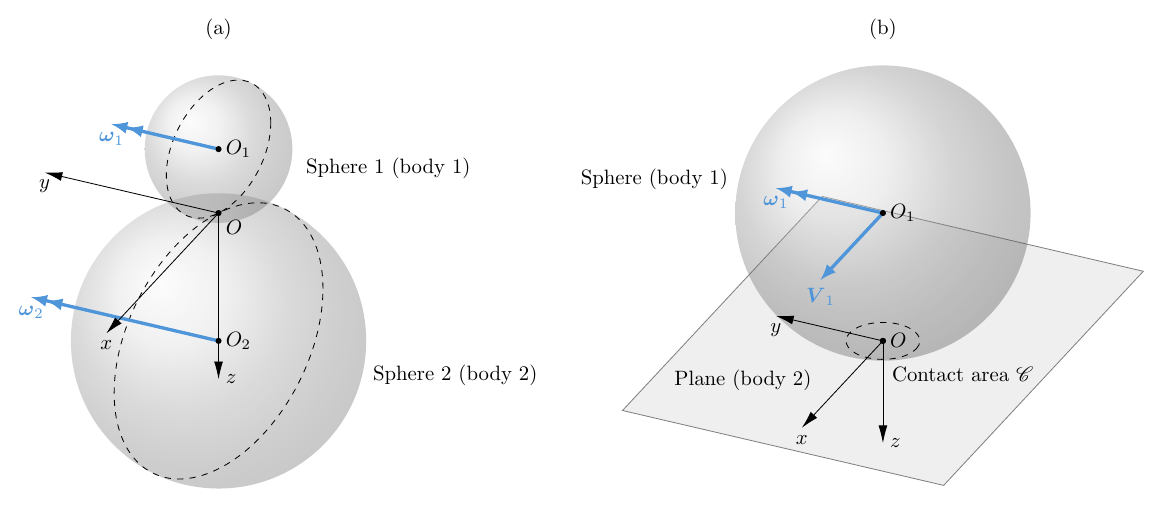} 
\caption{Rolling contact problem between: (a) two spheres with angular velocities $\bm{\omega}_1, \bm{\omega}_2 \in \mathbb{R}^3$; (b) a sphere translating and rolling over a stationary plane, where $\bm{V}_1\in \mathbb{R}^3$ denotes the translational velocity of its centre, and $\bm{\omega}_1\in \mathbb{R}^3$ its rolling velocity.}
\label{fig:RollingBodies}
\end{figure}

The following discussion is adapted from \cite{FrBDroll}. As illustrated in Fig.~\ref{fig:RollingBodies}, the rolling contact problem between two bodies is typically studied in a contact-fixed reference frame $(O;x,y,z)$, with the $x$-axis (longitudinal) oriented along the main rolling direction, the $z$-axis (vertical) pointing into one of the two bodies (often the lower one), and the $y$-axis (lateral) defined to complete a right-handed coordinate system. The origin $O$ coincides with the centroid of the (apparent), possibly time-varying, contact area $\mathscr{C}(t)$, which is often supposed to be independent of the tangential (longitudinal and lateral) interactions between the two bodies, and solely determined by the normal contact configuration \cite{KalkerBook}. Both the situations of reciprocal rolling, as depicted in Fig.~\ref{fig:RollingBodies}(a), and translational and rolling contact, as shown in Fig.~\ref{fig:RollingBodies}(b), may be considered: in the first case, the shape of the contact area may vary over time, but the origin $O$ is typically fixed; in the second scenario, the reference frame moves together with one of the two bodies (usually, the upper one).
In this paper, the contact area is assumed to be a regular closed, compact subset of $\mathbb{R}^2$, that is, $\mathscr{C}(t)\subset \mathbb{R}^2$, with boundary $\partial \mathscr{C}(t)$ and interior $\mathring{\mathscr{C}}(t)$. Inside the contact area, the displacement of a bristle, the force generated by each element, and the deformations will depend on their position, that is, $\bm{z}(t) = \bm{z}(\bm{x},t) = [z_x(\bm{x},t)\; z_y(\bm{x},t)]^{\mathrm{T}}$, $\bm{f}_i(t) = \bm{f}_i(\bm{x},t) = [f_{ix}(\bm{x},t)\; f_{iy}(\bm{x},t)]^{\mathrm{T}}$, and $\bm{z}_i(t) = \bm{z}_i(\bm{x},t) = [z_{ix}(\bm{x},t)\; z_{iy}(\bm{x},t)]^{\mathrm{T}}$, $i \in \{1,\dots,n\}$, with $\bm{x} \in \mathscr{C}(t)$. According to the Eulerian approach, the total time derivatives appearing in Eq.~\eqref{eq:ODEModel} become
\begin{subequations}\label{eq:Eulerian}
\begin{align}
\dot{\bm{z}}(\bm{x},t) = \dod{\bm{z}(\bm{x},t)}{t} &= \dpd{\bm{z}(\bm{x},t)}{t} + \bigl(\bm{V}(\bm{x},t)\cdot\nabla_{\bm{x}}\bigr)\bm{z}(\bm{x},t), \\
\dot{\bm{f}}_i(\bm{x},t) = \dod{\bm{f}_i(\bm{x},t)}{t} &= \dpd{\bm{f}_i(\bm{x},t)}{t} + \bigl(\bm{V}(\bm{x},t)\cdot\nabla_{\bm{x}}\bigr)\bm{f}_i(\bm{x},t), \\
\dot{\bm{z}}_i(\bm{x},t) = \dod{\bm{z}_i(\bm{x},t)}{t} &= \dpd{\bm{z}_i(\bm{x},t)}{t} + \bigl(\bm{V}(\bm{x},t)\cdot\nabla_{\bm{x}}\bigr)\bm{z}_i(\bm{x},t), \quad i \in \{1,\dots,n\},
\end{align}
\end{subequations}
where $\mathbb{R}^2 \ni \bm{V}(\bm{x},t) = [V_x(\bm{x},t) \; V_y(\bm{x},t)]^{\mathrm{T}}$ denotes the transport velocity, and $\mathbb{R}^2 \ni \nabla_{\bm{x}} \triangleq [\pd{}{x}\; \pd{}{y}]^{\mathrm{T}}$ is the tangential gradient. Moreover, considering the \emph{rolling speed} $[V\ped{min}, V\ped{max}] \ni V\ped{r}(t) \triangleq \norm{\bm{V}(\bm{0},t)}_2$, with $0 < V\ped{min} \leq V\ped{max}$, it is customary to define the \emph{travelled distance} $\mathbb{R}_{\geq 0} \ni s \triangleq \int_0^t V\ped{r}(t^\prime) \dif t^\prime$ to replace the time variable $t$. 

Utilising Eqs.~\eqref{eq:Eulerian}, the distributed FrBD$_{n+1}$-GM and FrBD$_{n+1}$-GKV models may be obtained directly from their lumped counterparts described by Eqs.~\eqref{eq:ODEModel} and~\eqref{eq:ODEModelKV}. Starting with the FrBD$_{n+1}$-GM formulation, introducing the state vector $\mathbb{R}^{2(n+1)} \ni \bm{u}(\bm{x},s) \triangleq [\bm{z}^{\mathrm{T}}(\bm{x},s) \; \bm{f}_1^{\mathrm{T}}(\bm{x},s) \; \bm{f}_2^{\mathrm{T}}(\bm{x},s) \; \dots\; \bm{f}_n^{\mathrm{T}}(\bm{x},s) ]^{\mathrm{T}}$, Eq.~\eqref{eq:ODEModel} may be recast as
\begin{subequations}\label{eq:PDenoMOdel09}
\begin{align}\label{eq:PDenoMOdel}
& \dpd{\bm{u}(\bm{x},s)}{s} + \bigl(\bar{\bm{V}}(\bm{x},s)\cdot\nabla_{\bm{x}}\bigr)\bm{u}(\bm{x},s) = \mathbf{\Sigma}\bigl(\bar{\bm{v}}\ped{r}(\bm{x},s),s\bigr)\bm{u}(\bm{x},s) + \bm{h}\bigl(\bar{\bm{v}}\ped{r}(\bm{x},s)\bigr), \quad \bm{x} \in \mathring{\mathscr{C}}(s), \; s \in (0,S), \\
& \bm{u}(\bm{x},0) = \bm{u}_0(\bm{x}), \quad \bm{x}\in \mathring{\mathscr{C}}_0,\label{eq:PDenoMOdelIC}
\end{align}
\end{subequations}
where $\mathring{\mathscr{C}}_0 \triangleq \mathring{\mathscr{C}}(0)$, $\mathbb{R}_{>0}\ni S \triangleq \int_0^T V\ped{r}(t)\dif t$, $\mathbb{R}^2 \ni \bar{\bm{V}}(\bm{x},s) = [\bar{V}_x(\bm{x},s)\; \bar{V}_y(\bm{x},s)]^{\mathrm{T}} \triangleq \bm{V}(\bm{x},s)/V\ped{r}(s)$, $\mathbb{R}^2 \ni \bar{\bm{v}}\ped{r}(\bm{x},s) = [\bar{v}_{\textnormal{r}x}(\bm{x},s)\; \bar{v}_{\textnormal{r}y}(\bm{x},s)]^{\mathrm{T}} \triangleq \bm{v}\ped{r}(\bm{x},s)/V\ped{r}(s)$, the functions $\mathbf{\Psi} : \mathbb{R}^2 \times \mathbb{R}_{\geq 0} \to \mathbf{M}_2(\mathbb{R})$ and $\bm{h}: \mathbb{R}^2 \to \mathbb{R}^{2(n+1)}$ given by 
\begin{subequations}\label{eq:Psifuns}
\begin{align}
\mathbf{\Psi}(\bar{\bm{v}}\ped{r},s) & \triangleq -\dfrac{1}{V\ped{r}(s)}\mathbf{M}^{-2}\bigl(V\ped{r}(s)\bar{\bm{v}}\ped{r}\bigr)\norm{\mathbf{M}\bigl(V\ped{r}(s)\bar{\bm{v}}\ped{r}\bigr)V\ped{r}(s)\bar{\bm{v}}\ped{r}}_{2,\varepsilon}, \label{eq:MPsi}\\
\bm{h}(\bar{\bm{v}}\ped{r})& \triangleq \mathbf{H}\bar{\bm{v}}\ped{r},
\end{align}
\end{subequations}
and $\mathbf{\Sigma} : \mathbb{R}^2\times\mathbb{R}_{\geq 0} \to \mathbf{M}_{2(n+1)}(\mathbb{R})$ and $\mathbf{H} \in \mathbf{M}_{2(n+1)\times 2}(\mathbb{R})$ read \begin{small}
\begin{subequations}\label{eq:Hfunct}
\begin{align}
\mathbf{\Sigma}(\bar{\bm{v}}\ped{r},s) & \triangleq \begin{bmatrix} \mathbf{\Psi}(\bar{\bm{v}}\ped{r},s)\bar{\mathbf{K}}_0 & \mathbf{\Psi}(\bar{\bm{v}}\ped{r},s) & \mathbf{\Psi}(\bar{\bm{v}}\ped{r},s) & \dots & \mathbf{\Psi}(\bar{\bm{v}}\ped{r},s) \\
 \bar{\mathbf{K}}_1 \mathbf{\Psi}(\bar{\bm{v}}\ped{r},s)\bar{\mathbf{K}}_0 & -\dfrac{\bm{\tau}_1^{-1}}{V\ped{r}(s)} + \bar{\mathbf{K}}_1\mathbf{\Psi}(\bar{\bm{v}}\ped{r},s) & \bar{\mathbf{K}}_1\mathbf{\Psi}(\bar{\bm{v}}\ped{r},s) & \dots & \bar{\mathbf{K}}_1 \mathbf{\Psi}(\bar{\bm{v}}\ped{r},s) \\
\bar{\mathbf{K}}_2 \mathbf{\Psi}(\bar{\bm{v}}\ped{r},s)\bar{\mathbf{K}}_0 & \bar{\mathbf{K}}_2\mathbf{\Psi}(\bar{\bm{v}}\ped{r},s) &  -\dfrac{\bm{\tau}_2^{-1}}{V\ped{r}(s)} + \bar{\mathbf{K}}_2\mathbf{\Psi}(\bar{\bm{v}}\ped{r},s)  &\dots & \bar{\mathbf{K}}_2 \mathbf{\Psi}(\bar{\bm{v}}\ped{r},s) \\
\vdots & \vdots & \vdots & \ddots & \vdots \\
\bar{\mathbf{K}}_n \mathbf{\Psi}(\bar{\bm{v}}\ped{r},s)\bar{\mathbf{K}}_0 & \bar{\mathbf{K}}_n\mathbf{\Psi}(\bar{\bm{v}}\ped{r},s) &   \bar{\mathbf{K}}_n\mathbf{\Psi}(\bar{\bm{v}}\ped{r},s)  &\dots & -\dfrac{\bm{\tau}_n^{-1}}{V\ped{r}(s)} + \bar{\mathbf{K}}_n \mathbf{\Psi}(\bar{\bm{v}}\ped{r},s)
 \end{bmatrix}, \\
\mathbf{H} & \triangleq -\begin{bmatrix} \mathbf{I}_2 & \bar{\mathbf{K}}_1 & \bar{\mathbf{K}}_2 & \dots & \bar{\mathbf{K}}_n \end{bmatrix}^{\mathrm{T}}.
\end{align}
\end{subequations}\end{small}
On the other hand, defining $\mathbb{R}^{2(n+1)} \ni \bm{u}(\bm{x},s) \triangleq [\bm{z}^{\mathrm{T}}(\bm{x},s) \; \bm{z}_1^{\mathrm{T}}(\bm{x},s) \; \bm{z}_2^{\mathrm{T}}(\bm{x},s) \; \dots\; \bm{z}_n^{\mathrm{T}}(\bm{x},s) ]^{\mathrm{T}}$, Eq.~\eqref{eq:ODEModelKV} admits a distributed parameter representation equivalent to~\eqref{eq:PDenoMOdel09}, with $\mathbf{\Psi}(\bar{\bm{v}}\ped{r},s)$ and $\bm{h}(\bar{\bm{v}}\ped{r})$ as in Eq.~\eqref{eq:Psifuns}, and
\begin{subequations}\label{eq:Hfunct2}
\begin{align}
\mathbf{\Sigma}(\bar{\bm{v}}\ped{r},s) & \triangleq \begin{bmatrix} \mathbf{\Psi}(\bar{\bm{v}}\ped{r},s)\bar{\mathbf{K}}_0 & -\mathbf{\Psi}(\bar{\bm{v}}\ped{r},s)\bar{\mathbf{K}}_0 & -\mathbf{\Psi}(\bar{\bm{v}}\ped{r},s)\bar{\mathbf{K}}_0 & \dots & -\mathbf{\Psi}(\bar{\bm{v}}\ped{r},s)\bar{\mathbf{K}}_0 \\
\dfrac{\bar{\mathbf{C}}_1^{-1}\bar{\mathbf{K}}_0}{V\ped{r}(s)} & -\dfrac{\bar{\mathbf{C}}_1^{-1}}{V\ped{r}(s)}(\bar{\mathbf{K}}_0+\bar{\mathbf{K}}_1) & -\dfrac{\bar{\mathbf{C}}_1^{-1}\bar{\mathbf{K}}_0}{V\ped{r}(s)} & \dots & -\dfrac{\bar{\mathbf{C}}_1^{-1}\bar{\mathbf{K}}_0}{V\ped{r}(s)} \\
\dfrac{\bar{\mathbf{C}}_2^{-1}\bar{\mathbf{K}}_0}{V\ped{r}(s)} & -\dfrac{\bar{\mathbf{C}}_2^{-1}\bar{\mathbf{K}}_0}{V\ped{r}(s)} &  -\dfrac{\bar{\mathbf{C}}_2^{-1}}{V\ped{r}(s)}(\bar{\mathbf{K}}_0 + \bar{\mathbf{K}}_2) &\dots & -\dfrac{\bar{\mathbf{C}}_2^{-1}\bar{\mathbf{K}}_0}{V\ped{r}(s)} \\
\vdots & \vdots & \vdots & \ddots & \vdots \\
\dfrac{\bar{\mathbf{C}}_n^{-1}\bar{\mathbf{K}}_0}{V\ped{r}(s)} & -\dfrac{\bar{\mathbf{C}}_n^{-1}\bar{\mathbf{K}}_0}{V\ped{r}(s)} &   -\dfrac{\bar{\mathbf{C}}_n^{-1}\bar{\mathbf{K}}_0}{V\ped{r}(s)}  &\dots & -\dfrac{\bar{\mathbf{C}}_n^{-1}}{V\ped{r}(s)}(\bar{\mathbf{K}}_0 + \bar{\mathbf{K}}_n)
 \end{bmatrix},  \\
\mathbf{H} & \triangleq -\begin{bmatrix} \mathbf{I}_2 & \mathbf{0} & \mathbf{0} & \dots & \mathbf{0}\end{bmatrix}^{\mathrm{T}}.
\end{align}
\end{subequations}
Equations~\eqref{eq:PDenoMOdel09}-\eqref{eq:Hfunct2} define systems of first-order hyperbolic PDEs with internal relaxation dynamics, whose characteristic structure directly reflects the rolling kinematics of the contact interface.
In particular, to ensure well-posedness, Eq.~\eqref{eq:PDenoMOdel09} needs to be supplemented with a set of BCs, as discussed next in Sect.~\ref{sect:BC}.

\subsubsection{Boundary condition (BC)}\label{sect:BC}
Appropriate BC for the PDE~\eqref{eq:PDenoMOdel} may be formulated by defining the \emph{leading edge} $\mathscr{L}(s)$, \emph{neutral edge} $\mathscr{N}(s)$, and \emph{trailing edge} $\mathscr{T}(s)$ as follows \cite{FrBDroll}:
\begin{subequations}\label{eq:Inflow0Outflow}
\begin{align}
\mathscr{L}(s) &\triangleq \partial \mathscr{C}_{-}(s)= \Bigl\{\bm{x}\in \partial \mathscr{C}(s) \mathrel{\Big|}\bigl[\bar{\bm{V}}(\bm{x},s)-\bar{\bm{V}}_{\partial \mathscr{C}}(\bm{x},s)\bigr] \cdot \hat{\bm{n}}_{\partial \mathscr{C}}(\bm{x},s) < 0\Bigr\}, \\
\mathscr{N}(s) &\triangleq \partial \mathscr{C}_{0}(s) = \Bigl\{\bm{x}\in \partial \mathscr{C}(s) \mathrel{\Big|} \bigl[\bar{\bm{V}}(\bm{x},s)-\bar{\bm{V}}_{\partial \mathscr{C}}(\bm{x},s)\bigr] \cdot \hat{\bm{n}}_{\partial \mathscr{C}}(\bm{x},s) = 0\Bigr\}, \\
\mathscr{T}(s) &\triangleq \partial \mathscr{C}_{+}(s)= \Bigl\{\bm{x}\in \partial \mathscr{C}(s) \mathrel{\Big|} \bigl[\bar{\bm{V}}(\bm{x},s)-\bar{\bm{V}}_{\partial \mathscr{C}}(\bm{x},s)\bigr] \cdot \hat{\bm{n}}_{\partial \mathscr{C}}(\bm{x},s) > 0\Bigr\}, 
\end{align}
\end{subequations}
where $\hat{\bm{n}}_{\partial \mathscr{C}}(\bm{x},s)\in \mathbb{R}^2$ denotes the outward unit normal to $\partial \mathscr{C}(s)$, and $\bar{\bm{V}}_{\partial \mathscr{C}}(\bm{x},s) \in \mathbb{R}^2$ is its nondimensional velocity. 
For a transport equation like Eq.~\eqref{eq:PDenoMOdel}, the natural BC should be prescribed at the inflow boundary, that is, on the leading edge $\mathscr{L}(s)$, for all $s \in (0,S)$. In particular, enforcing $\bm{u}(\bm{x},s) = \bm{0}$ ensures the continuity of both the deformations and generated forces at the transition between the free portions of the bodies and those making contact. This resolves the long-standing issues arising from lower-order models, as those presented in \cite{FrBDroll}, which only guarantee the continuity of the deformations \cite{TsiotrasConf,Tsiotras1,Tsiotras3,Deur0,Deur1,Deur2}. 

\subsubsection{Tangential forces and vertical moment}\label{ect:forces}
Recalling Eqs.~\eqref{eq:force00} and~\eqref{eq:force00GKV}, the total force generated by the bristle deflection may be generally expressed as
\begin{align}\label{eq:ftot}
\bm{f}(\bm{x},s) = \mathbf{C}\bm{u}(\bm{x},s),
\end{align}
with the matrix $\mathbf{C} \in \mathbf{M}_{2\times 2(n+1)}$ reading
\begin{align}\label{eq:C1}
\mathbf{C} \triangleq \begin{bmatrix}\bar{\mathbf{K}}_0 & \mathbf{I}_2 & \mathbf{I}_2 & \dots \mathbf{I}_2\end{bmatrix},
\end{align}
for the FrBD$_{n+1}$-GM model, and 
\begin{align}\label{eq:C2}
\mathbf{C} \triangleq \begin{bmatrix}\bar{\mathbf{K}}_0 & -\bar{\mathbf{K}}_0 & -\bar{\mathbf{K}}_0 & \dots & -\bar{\mathbf{K}}_0\end{bmatrix},
\end{align}
for the FrBD$_{n+1}$-GKV one.
Starting with Eqs.~\eqref{eq:ftot},~\eqref{eq:C1}, and~\eqref{eq:C2}, the tangential forces $\mathbb{R}\ni \bm{F}_{\bm{x}}(s) = [F_x(s)\; F_y(s)]^{\mathrm{T}}$ and the vertical moment $M_z(s)\in \mathbb{R}$ may then be computed as
\begin{subequations}\label{eq:FandM}
\begin{align}
\bm{F}_{\bm{x}}(s) & = \iint_{\mathscr{C}(s)}p(\bm{x},s)\bm{f}(\bm{x},s) \dif \bm{x}, \label{eq:Fundef}\\
M_z(s) & = \iint_{\mathscr{C}(s)}p(\bm{x},s)\bigl[ xf_y(\bm{x},s)-yf_x(\bm{x},s)\bigr] \dif \bm{x}, \quad s \in [0,S],\label{eq:Mzunderfr}
\end{align}
\end{subequations}
where $p \in C^0(\mathscr{C}\times[0,S];\mathbb{R}_{\geq 0})$ indicates the pressure distribution inside the contact area. 

In alternative to Eq.~\eqref{eq:Mzunderfr}, the vertical moment can also be evaluated considering the deformed configuration:
\begin{align}\label{eq:Malt}
M_z(s) & = \iint_{\mathscr{C}(s)}p(\bm{x},s)\Bigl[\bigl(x+z_x(\bm{x},s)\bigr)f_y(\bm{x},s)-\bigl(y+z_y(\bm{x},s)\bigr)f_x(\bm{x},s)\Bigr] \dif \bm{x}, \quad s \in [0,S].
\end{align}
Equation~\eqref{eq:Fundef} remains formally unchanged.

The next Sect.~\ref{sect:modelRollSimple} presents three different FrBD$_{n+1}$-GM rolling contact models.

\subsection{FrBD$_{n+1}$ rolling contact models}\label{sect:modelRollSimple}
In order to solve the PDE~\eqref{eq:PDenoMOdel09}, it is necessary to specify the (nondimensional) transport and rigid relative velocities $\bar{\bm{V}}(\bm{x},s)$ and $\bar{\bm{v}}\ped{r}(\bm{x},s)$. Concerning the former, for a viscoelastic body rolling and sliding over a flat substrate, the most general expression may be deduced as \cite{FrBDroll}
\begin{align}\label{eq:generalV}
\bar{\bm{V}}(\bm{x},s) = -\begin{bmatrix} \varepsilon_y(s) \\ -\varepsilon_x(s) \end{bmatrix} + \mathbf{A}_{\varphi_1}(s)\bm{x},
\end{align}
where $\mathbb{R}^2 \ni \bm{\varepsilon}(s) = [\varepsilon_x(s)\; \varepsilon_y(s)]^{\mathrm{T}}$, with $\norm{\bm{\varepsilon}(s)}_2 = 1$, and $\mathbf{A}_{\varphi_1}(s) \in \mathbf{M}_2(\mathbb{R})$ reads
\begin{align}\label{eq:geometricSPinTensor}
\mathbf{A}_{\varphi_1}(s) \triangleq  \begin{bmatrix}0 & \varphi_1(s) \\ -\varphi_1(s) & 0 \end{bmatrix},
\end{align}
being $\varphi_1(s) \in \mathbb{R}$ a spin component depending on the reciprocal deformative behaviour between the contacting bodies \cite{FrBDroll}. 
On the other hand, the nondimensional rigid relative velocity is generally a function of the bristle deformation \cite{FrBDroll}, according to
\begin{align}\label{eq:RigidVelSpin}
\bar{\bm{v}}\ped{r}\bigl(\bm{u}(\bm{x},s),\bm{x},s\bigr) =\bar{\bm{v}}(\bm{x},s) - \mathbf{A}_{\varphi_2}(s)\bm{z}(\bm{x},s),
\end{align}
with $\mathbb{R}^2 \ni \bar{\bm{v}}(\bm{x},s) = [\bar{v}_x(\bm{x},s) \; \bar{v}_y(\bm{x},s)]^{\mathrm{T}}$ given by
\begin{align}\label{eq:v0}
\bar{\bm{v}}(\bm{x},s) = -\bm{\sigma}(s) - \mathbf{A}_\varphi(s)\bm{x},
\end{align}
where $\mathbb{R}^2 \ni \bm{\sigma}(s) = [\sigma_x(s) \; \sigma_y(s)]^{\mathrm{T}}$ denotes the \emph{translational slip}, and the matrix $\mathbf{A}_{\varphi}(s) \in \mathbf{M}_2(\mathbb{R})$ reads
\begin{align}\label{eq:Aphi}
\mathbf{A}_\varphi (s)&  \triangleq \begin{bmatrix} 0 & -\varphi (s)\\
\varphi(s) & 0 \end{bmatrix},
\end{align}
being $\varphi(s) \in \mathbb{R}$ the \emph{total spin slip}, which is the sum of the \emph{geometrical spin} $\varphi_\gamma (s) \in \mathbb{R}$, and the \emph{effective spin} $\varphi_\psi(s)$:
\begin{align}
\varphi(s) = \varphi_\gamma(s) + \varphi_\psi(s).
\end{align}
Finally, the matrix $\mathbf{A}_{\varphi_2}(s) \in \mathbf{M}_2(\mathbb{R})$ in Eq.~\eqref{eq:RigidVelSpin} writes
\begin{align}\label{eq:turnSpinMatrix}
\mathbf{A}_{\varphi_2}(s) & \triangleq \begin{bmatrix} 0 & -\varphi_2(s) \\ \varphi_2(s) & 0\end{bmatrix},
\end{align}
where $\varphi_2(s) \in \mathbb{R}$ is a second spin component, whose definition also depends on the reciprocal material properties.  For a general interpretation of the terms $\varphi_1(s)$ and $\varphi_2(s)$, the reader is referred to \cite{FrBDroll}. For notational convenience, in this paper, they are collected into the vector $\mathbb{R}^2 \ni \bm{\varphi}(s) = [\varphi_1(s) \; \varphi_2(s)]^{\mathrm{T}}$.

Rolling contact models with different orders of complexity may be deduced from Eqs.~\eqref{eq:generalV}-\eqref{eq:turnSpinMatrix} by further simplifying the expressions for the transport and rigid relative velocities. Specifically, by approximating $\bar{\bm{V}}(\bm{x},s) \approx -[1\; 0]^{\mathrm{T}}$ in Eq.~\eqref{eq:generalV} and neglecting the term $\mathbf{A}_{\varphi_2(s)}\bm{z}(\bm{x},s)$ in~\eqref{eq:RigidVelSpin}, the \emph{standard linear FrBD$_{n+1}$ rolling contact models} formalised below are obtained (\modref{stdmodel}).
\begin{modelbox}[stdmodel]{Standard linear FrBD$_{n+1}$ models}
\begin{subequations}\label{eq:standard0}
\begin{align}
\begin{split}
& \dpd{\bm{u}(\bm{x},s)}{s} -\dpd{\bm{u}(\bm{x},s)}{x} =\mathbf{\Sigma}\bigl(\bar{\bm{v}}(\bm{x},s),s\bigr)\bm{z}(\bm{x},s) + \bm{h}\bigl(\bar{\bm{v}}(\bm{x},s)\bigr), \quad \bm{x} \in \mathring{\mathscr{C}}(s), \; s \in (0,S),
\end{split}\\
& \bm{u}(\bm{x},s) = \bm{0}, \quad \bm{x}\in \mathscr{L}(s), \;  s \in (0,S), \\
& \bm{u}(\bm{x},0) = \bm{u}_0(\bm{x}), \quad \bm{x}\in \mathring{\mathscr{C}}_0.
\end{align}
\end{subequations}
\end{modelbox}
\modref{stdmodel} can adequately describe rolling contact processes evolving along a main direction in the presence of small spin slips, such as wheel-rail and tyre-road interactions in normal driving conditions (small camber angles and turn spins). Considering instead the exact expressions Eqs.~\eqref{eq:generalV}-\eqref{eq:turnSpinMatrix}, the \emph{semilinear FrBD$_{n+1}$ models for large spin slips} (\modref{semilinmodel}) are formulated below.
\begin{modelbox}[semilinmodel]{Semilinear FrBD$_{n+1}$ models for large spin slips}
\begin{subequations}
\begin{align}\label{eq:semilinear}
\begin{split}
& \dpd{\bm{u}(\bm{x},s)}{s} + \bigl(\bar{\bm{V}}(\bm{x},s)\cdot \nabla_{\bm{x}}\bigr)\bm{u}(\bm{x},s) =\mathbf{\Sigma}\Bigl(\bar{\bm{v}}\ped{r}\bigl(\bm{u}(\bm{x},s),\bm{x},s\bigr),s\Bigr)\bm{u}(\bm{x},s)\\
& \qquad \qquad \qquad \qquad \qquad \qquad \qquad\qquad + \bm{h}\Bigl(\bar{\bm{v}}\ped{r}\bigl(\bm{u}(\bm{x},s),\bm{x},s\bigr)\Bigr), \quad \bm{x} \in \mathring{\mathscr{C}}(s), \; s \in (0,S),
\end{split}\\
& \bm{u}(\bm{x},s) = \bm{0}, \quad \bm{x}\in \mathscr{L}(s), \;  s \in (0,S), \\
& \bm{u}(\bm{x},0) = \bm{u}_0(\bm{x}), \quad \bm{x}\in \mathring{\mathscr{C}}_0.
\end{align}
\end{subequations}
\end{modelbox}
Compared to~\modref{stdmodel},~\modref{semilinmodel} capture nonlinear effects induced by large spin slips, which arise, for instance, in automotive tyres subjected to high camber angles or during low-speed parking manoeuvres. However, in contrast to~\modref{stdmodel}, they are highly nonlinear in the variable $\bm{u}(\bm{x},s)$, which makes their numerical resolution more delicate. 

Finally, an approximation to the semilinear PDE~\eqref{eq:semilinear} may be recovered by replacing $\bar{\bm{v}}\ped{r}(\bm{z}(\bm{x},s),\bm{x},s)$ by $\bar{\bm{v}}(\bm{x},s)$ in the nonlinear function $\mathbf{\Sigma}(\cdot,\cdot)$, yielding the following \emph{linear FrBD$_{n+1}$ models for large spin slips} (\modref{linmodel2}).
\begin{modelbox}[linmodel2]{Linear FrBD$_{n+1}$ models for large spin slips}
\begin{subequations}\label{eq:linwasdir}
\begin{align}\label{eq:linwasdir2}
\begin{split}
& \dpd{\bm{u}(\bm{x},s)}{s} + \bigl(\bar{\bm{V}}(\bm{x},s)\cdot \nabla_{\bm{x}}\bigr)\bm{u}(\bm{x},s) =\mathbf{\Sigma}\bigl(\bar{\bm{v}}(\bm{x},s),s\bigr)\bm{u}(\bm{x},s) \\
&\qquad \qquad \qquad \qquad \qquad \qquad \qquad\qquad + \bm{h}\Bigl(\bar{\bm{v}}\ped{r}\bigl(\bm{u}(\bm{x},s),\bm{x},s\bigr)\Bigr), \quad \bm{x} \in \mathring{\mathscr{C}}(s), \; s \in (0,S),
\end{split}\\
& \bm{u}(\bm{x},s) = \bm{0}, \quad \bm{x}\in \mathscr{L}(s), \;  s \in (0,S), \\
& \bm{u}(\bm{x},0) = \bm{u}_0(\bm{x}), \quad \bm{x}\in \mathring{\mathscr{C}}_0.
\end{align}
\end{subequations}
\end{modelbox}
\modref{linmodel2} are appropriate in regimes where the spin $\varphi_1(s)$ may attain large values, whereas $\varphi_2(s)$ remains moderate. They may be regarded as a linear, zeroth-order approximation to~\modref{semilinmodel}, and can therefore be employed as an initial iterate in fixed-point-based numerical solution procedures.

\modref{stdmodel},~\ref{semilinmodel}, and~\ref{linmodel2} represent the FrBD$_{n+1}$-GM and FrBD$_{n+1}$-GKV counterparts of the FrBD$_1$-KV formulations introduced in \cite{FrBDroll}. The next Sect.~\ref{sect:math} proceeds to the rigorous analysis of~\modref{stdmodel} and~\ref{linmodel2} under standard assumptions.

\section{Mathematical properties}\label{sect:math}
The present section is devoted to the mathematical analysis of~\modref{stdmodel} and~\ref{linmodel2}. More specifically, Sect.~\ref{sect:wellp} investigates their well-posedness, whereas passivity is studied in Sect.~\ref{sect:pass}.

\subsection{Well-posedness}\label{sect:wellp}
Well-posedness for~\modref{stdmodel} and~\ref{linmodel2} may be asserted within different functional settings. For simplicity, the following results are enounced considering a time-invariant contact area, which permits recovering existence and uniqueness results by invoking standard semigroup arguments \cite{Pazy,Tanabe1,Tanabe}. Extensions to slowly varying domains may be pursued using time-dependent semigroups \cite{FrBDroll} or ALE-based techniques.

For what follows, it is convenient to introduce the functions
\begin{subequations}\label{eq:Sigmaf0}
\begin{align}
\tilde{\bm{h}}(\bm{x},s) & \triangleq  \bm{h}\bigl(\bar{\bm{v}}(\bm{x},s)\bigr), \\
\tilde{\mathbf{\Sigma}}(\bm{x},s) & \triangleq \mathbf{\Sigma}\bigl(\bar{\bm{v}}(\bm{x},s),s\bigr), \\
\tilde{\mathbf{\Sigma}}_{\varphi}(\bm{x},s) & \triangleq \mathbf{\Sigma}\bigl(\bar{\bm{v}}(\bm{x},s),s\bigr)- \mathbf{H}\mathbf{A}_{\varphi_2}(s).
\end{align}
\end{subequations}
so that Eq.~\eqref{eq:standard0} may be restated as
\begin{subequations}\label{eq:standard2}
\begin{align}
\begin{split}
& \dpd{\bm{u}(\bm{x},s)}{s} -\dpd{\bm{u}(\bm{x},s)}{x} =\tilde{\mathbf{\Sigma}}(\bm{x},s)\bm{u}(\bm{x},s) + \tilde{\bm{h}}(\bm{x},s), \quad (\bm{x},s) \in \mathring{\mathscr{C}}\times (0,S),\label{eq:standard2PDE}
\end{split}\\
& \bm{u}(\bm{x},s) = \bm{0}, \quad (\bm{x},s) \in \mathscr{L}\times (0,S), \label{eq:BCquasiStandard}\\
& \bm{u}(\bm{x},0) = \bm{u}_0(\bm{x}), \quad \bm{x}\in \mathring{\mathscr{C}}.
\end{align}
\end{subequations}
For the simplified Eq.~\eqref{eq:standard2}, existence and uniqueness of the solution are asserted below by Theorem~\ref{thm:ex1}. 
\begin{theorem}[Existence and uniqueness of solutions of~\modref{stdmodel}]\label{thm:ex1}
Suppose that $\mathring{\mathscr{C}}\subset \mathbb{R}^2$ is bounded, with boundary $\partial \mathscr{C}$ piecewise $C^1$. Then, for all $\tilde{\mathbf{\Sigma}} \in C^0(\mathscr{C}\times[0,S];\mathbf{M}_{2(n+1)}(\mathbb{R}))$ and $\tilde{\bm{h}} \in C^0([0,S];L^2(\mathring{\mathscr{C}};\mathbb{R}^{2(n+1)}))$ as in Eq.~\eqref{eq:Sigmaf0}, and ICs $\bm{u}_0 \in L^2(\mathring{\mathscr{C}};\mathbb{R}^{2(n+1)})$, the PDE~\eqref{eq:standard2} admits a unique \emph{mild solution} $\bm{u} \in C^0([0,S];L^2(\mathring{\mathscr{C}};\mathbb{R}^{2(n+1)}))$. Additionally, if $\tilde{\mathbf{\Sigma}} \in C^1(\mathscr{C}\times[0,S];\mathbf{M}_{2(n+1)}(\mathbb{R}))$, $\tilde{\bm{h}} \in C^1([0,S];L^2(\mathring{\mathscr{C}};\mathbb{R}^{2(n+1)}))$, and the IC $\bm{u}_0 \in \mathscr{D}(\mathscr{A})$, with $\mathscr{D}(\mathscr{A}) \triangleq \{\bm{v} \in L^2(\mathring{\mathscr{C}};\mathbb{R}^{2(n+1)}) \mathrel{|} \pd{\bm{v}}{x} \in  L^2(\mathring{\mathscr{C}};\mathbb{R}^{2(n+1)}), \; \eval[0]{\bm{v}}_{\mathscr{L}} = \bm{0}\}$, the solution is \emph{classical}, that is, $\bm{u} \in C^1([0,S];\allowbreak L^2(\mathring{\mathscr{C}};\mathbb{R}^{2(n+1)})) \cap C^0([0,S];\mathscr{D}(\mathscr{A}))$.
\begin{proof}[Proof]
The result follows along the same lines as the proof of Theorem 3.1 in \cite{FrBDroll}.
\end{proof}
\end{theorem}
Owing to these premises, the PDE~\eqref{eq:linwasdir} is reformulated as
\begin{subequations}\label{eq:quasiStationary0}
\begin{align}
\begin{split}
& \dpd{\bm{u}(\bm{x},s)}{s} + \bigl(\bar{\bm{V}}(\bm{x})\cdot \nabla_{\bm{x}}\bigr)\bm{u}(\bm{x},s) =\tilde{\mathbf{\Sigma}}_{\varphi}(\bm{x},s)\bm{u}(\bm{x},s) +\tilde{\bm{h}}(\bm{x},s),\quad (\bm{x},s) \in\mathring{\mathscr{C}} \times (0,S),\label{eq:quasiStationary0dy}
\end{split}\\
& \bm{u}(\bm{x},s) = \bm{0}, \quad (\bm{x},s) \in \mathscr{L}\times (0,S),\label{eq:BCquasistesd} \\
& \bm{u}(\bm{x},0) = \bm{u}_0(\bm{x}), \quad \bm{x} \in \mathring{\mathscr{C}},
\end{align}
\end{subequations}
with the nondimensional transport velocity $\bar{\bm{V}}(\bm{x})$ given, in the most general case, by
\begin{align}\label{eq:Vss}
\bar{\bm{V}}(\bm{x}) = -\begin{bmatrix} \varepsilon_y \\ - \varepsilon_x\end{bmatrix} + \mathbf{A}_{\varphi_1}\bm{x}.
\end{align}
Accordingly, well-posedness results are enounced in Theorem~\ref{thm:ex2} below.
\begin{theorem}[Existence and uniqueness of solutions of~\modref{linmodel2}]\label{thm:ex2}
Suppose that $\bar{\bm{V}} \in C^1(\mathscr{C};\mathbb{R}^2)$ reads as in Eq.~\eqref{eq:Vss}, and that $\mathring{\mathscr{C}}\subset \mathbb{R}^2$ is bounded, with boundary $\partial \mathscr{C}$ piecewise $C^1$. Then, for all $\tilde{\mathbf{\Sigma}}_{\varphi} \in C^0(\mathscr{C}\times[0,S];\mathbf{M}_{2(n+1)}(\mathbb{R}))$ and $\tilde{\bm{h}} \in C^0([0,S];L^2(\mathring{\mathscr{C}};\mathbb{R}^{2(n+1)}))$ as in Eq.~\eqref{eq:Sigmaf0}, and ICs $\bm{u}_0 \in L^2(\mathring{\mathscr{C}};\mathbb{R}^{2(n+1)})$, the PDE~\eqref{eq:quasiStationary0} admits a unique mild solution $\bm{u} \in C^0([0,S];L^2(\mathring{\mathscr{C}};\mathbb{R}^{2(n+1)}))$. Additionally, if $\tilde{\mathbf{\Sigma}}_{\varphi} \in C^1(\mathscr{C}\times[0,S];\mathbf{M}_{2(n+1)}(\mathbb{R}))$, $\tilde{\bm{h}} \in C^1([0,S];L^2(\mathring{\mathscr{C}};\mathbb{R}^{2(n+1)}))$, and the IC $\bm{u}_0 \in \mathscr{D}(\mathscr{A})$, with $\mathscr{D}(\mathscr{A}) \triangleq \{\bm{v} \in L^2(\mathring{\mathscr{C}};\mathbb{R}^{2(n+1)}) \mathrel{|} (\bar{\bm{V}}\cdot\nabla_{\bm{x}})\bm{v} \in  L^2(\mathring{\mathscr{C}};\mathbb{R}^{2(n+1)}), \; \eval[0]{\bm{v}}_{\mathscr{L}} = \bm{0}\}$, the solution is classical, that is, $\bm{u} \in C^1([0,S];L^2(\mathring{\mathscr{C}};\mathbb{R}^{2(n+1)})) \cap C^0([0,S];\mathscr{D}(\mathscr{A}))$.
\begin{proof}
The result follows along the same lines as the proof of Theorem 3.2 in \cite{FrBDroll}.
\end{proof}
\end{theorem}
Some comments about the regularity of the original matrices and inputs are as follows. Starting with~\modref{stdmodel}, $\mathbf{M}\in C^0(\mathbb{R}^2; \mathbf{Sym}_2(\mathbb{R}))$ ensures the existence of mild solutions for $\bar{\bm{v}} \in C^0(\mathscr{C}\times[0,S];\mathbb{R}^2)$ and $V\ped{r} \in C^0([0,S]; [V\ped{min}, V\ped{max}])$. Additionally, by composition of continuously differentiable functions, $\mathbf{M}\in C^1(\mathbb{R}^2; \mathbf{Sym}_2(\mathbb{R}))$ implies $\tilde{\mathbf{\Sigma}} \in C^1(\mathscr{C}\times[0,S];\mathbf{M}_2(\mathbb{R}))$ and $\tilde{\bm{h}} \in C^1([0,S];L^2(\mathring{\mathscr{C}};\mathbb{R}^2))$ for all $\bar{\bm{v}} \in C^1(\mathscr{C}\times[0,S];\mathbb{R}^2)$ and $V\ped{r} \in C^1([0,S]; [V\ped{min}, V\ped{max}])$. Essentially, for the regularised version (with $\varepsilon \in \mathbb{R}_{>0}$), classical solutions are guaranteed for sufficiently smooth slip and rolling velocities. Theorem~\ref{thm:ex2} demands additionally $\bm{\varphi} \in \mathbb{R}\times C^0([0,S];\mathbb{R})$ for mild solutions, and $\bm{\varphi} \in \mathbb{R}\times C^1([0,S];\mathbb{R})$ (with regularisation parameter $\varepsilon \in \mathbb{R}_{>0}$) for classical solutions. 

For~\modref{stdmodel} and~\ref{linmodel2}, ISS and IOS may also be proved for all inputs $\bar{\bm{v}} \in C^1(\mathscr{C}\times\mathbb{R}_{\geq 0};\mathbb{R}^2) \cap L^\infty(\mathscr{C}\times\mathbb{R}_{\geq 0};\mathbb{R}^2)$, rolling speeds $V\ped{r} \in C^1(\mathbb{R}_{\geq 0};[V\ped{min},V\ped{max}])$, spins $\bm{\varphi} \in \mathbb{R}\times C^1(\mathbb{R}_{\geq 0};\mathbb{R})\cap \mathbb{R}\times L^\infty(\mathbb{R}_{\geq 0};\mathbb{R})$, and ICs $\bm{u}_0 \in \mathscr{D}(\mathscr{A})$ under the same assumptions as Lemma 4.1 and Corollary 4.1 in \cite{FrBDroll}.

\subsection{Passivity}\label{sect:pass}
Dissipativity and passivity arguably constitute the most important mathematical properties of a \emph{bona fide} friction model, as they ensure consistency with the inherently dissipative nature of the underlying physical phenomenon. For the rolling contact processes described by Eqs.~\eqref{eq:standard2} and~\eqref{eq:quasiStationary0}, dissipativity expresses the fact that friction leads to a net energy loss within the contact region, whereas passivity ensures that no energy is generated during frictional rolling contact. For the FrBD$_{n+1}$ models presented in this paper, passivity is also particularly relevant for control-oriented simulations, as it guarantees stability when coupled with mechanical systems or observers. It is worth emphasising that, in this context, passivity should be understood as a property of the process rather than of the friction model itself.
Starting with dissipativity, observing that
\begin{align}
\bm{v}\ped{s}(\bm{x},s) = V\ped{r}(s)\dod{\bm{z}(\bm{x},s)}{s} + V\ped{r}(s)\bar{\bm{v}}\ped{r}(\bm{x},s), \quad s \in [0,S],
\end{align}
and in analogy to what done in Sect.~\ref{sect:DynamicDer}, for $p \in C^0(\mathscr{C}\times [0,S];\mathbb{R}_{\geq 0})$ and $V\ped{r}\in C^0([0,S];[V\ped{min},V\ped{max}])$, the dissipated power may be inferred to be
\begin{align}\label{eq:frictionPass}
\begin{split}
-\bigl\langle p(\cdot,s)\bm{f}(\cdot,s),\bm{v}\ped{s}(\cdot,s)\bigr\rangle_{L^2(\mathring{\mathscr{C}};\mathbb{R}^2)} & = -\iint_{\mathscr{C}} p(\bm{x},s)\bm{f}^{\mathrm{T}}(\bm{x},s)\bm{v}\ped{s}(\bm{x},s) \dif \bm{x} \\
& =- \iint_{\mathscr{C}} V\ped{r}(s)p(\bm{x},s)\bm{f}^{\mathrm{T}}(\bm{x},s)\mathbf{\Psi}\bigl(\bar{\bm{v}}\ped{r}(\bm{x},s),s\bigr)\bm{f}(\bm{x},s) \dif \bm{x} \geq 0, \quad s \in [0,S],
\end{split}
\end{align}
which is consistent with the nature of the friction process. Equation~\eqref{eq:frictionPass} evaluates the dissipation rate considering the physically correct velocity, namely the sliding velocity of the tip of the bristles. However, from a dynamical system theory perspective, the input-output behaviour of the PDEs~\eqref{eq:standard2} and~\eqref{eq:quasiStationary0} may also be studied regarding the velocity $\bm{v}(\bm{x},s)$ (or equivalently $\bar{\bm{v}}(\bm{x},s)$) as the input. Owing to these preliminary considerations, the mathematical definitions of dissipativity and passivity, adopted from \cite{FrBDroll}, are given below.
\begin{definition}[Dissipativity and passivity]\label{def:idss2}
The system described by the PDE~\eqref{eq:standard2} (resp.~\eqref{eq:quasiStationary0}) with output~\eqref{eq:Fundef} (resp.~\eqref{eq:Fundef} and~\eqref{eq:Malt}) is called \emph{dissipative} if, for all inputs $\bar{\bm{v}} \in C^1(\mathscr{C}\times\mathbb{R}_{\geq 0};\mathbb{R}^2) \cap L^\infty(\mathscr{C}\times\mathbb{R}_{\geq 0};\mathbb{R}^2)$, rolling speeds $V\ped{r} \in [V\ped{min},V\ped{max}]$, spins $\bm{\varphi} \in \mathbb{R}\times \{0\}$, and ICs $\bm{u}_0 \in \mathscr{D}(\mathscr{A})$, there exist a supply rate $w : L^2(\mathring{\mathscr{C}};\mathbb{R}^2)\times L^2(\mathring{\mathscr{C}};\mathbb{R}^2) \to \mathbb{R}$ and a storage function $W : L^2(\mathring{\mathscr{C}};\mathbb{R}^{2(n+1)}) \to \mathbb{R}_{\geq 0}$ such that 
\begin{align}\label{eq:Fvres}
\begin{split}
&\int_0^s w\bigl(\bm{f}(\cdot,s^\prime),-\bar{\bm{v}}\ped{r}(\cdot,s^\prime)\bigr) \dif s^\prime = \int_0^s w\bigl(\bm{f}(\cdot,s^\prime),-\bar{\bm{v}}(\cdot,s^\prime)\bigr) \dif s^\prime \geq W\bigl(\bm{u}(\cdot,s)\bigr)-W\bigl(\bm{u}_0(\cdot)\bigr), \quad s \in [0,S].
\end{split}
\end{align}
It is called \emph{passive} if $w(\bm{f}(\cdot,s),-\bar{\bm{v}}(\cdot,s)) =-\langle p(\cdot)\bm{f}(\cdot,s), \bar{\bm{v}}\ped{r}(\cdot,s)\rangle_{L^2(\mathring{\mathscr{C}};\mathbb{R}^2)}= -\langle p(\cdot)\bm{f}(\cdot,s), \bar{\bm{v}}(\cdot,s)\rangle_{L^2(\mathring{\mathscr{C}};\mathbb{R}^2)}$.
\end{definition}
The two main results of this section are formalised in Lemmata~\ref{lemma:DissF20} and~\ref{lemma:DissF202}, which consider respectively the FrBD$_{n+1}$-GM and FrBD$_{n+1}$-GKV realisations of the distributed friction model.

\begin{lemma}[Passivity of the FrBD$_{n+1}$-GM models]\label{lemma:DissF20}
Consider the PDE~\eqref{eq:standard2} (resp.~\eqref{eq:quasiStationary0}) with $\mathbf{\Sigma}(\bar{\bm{v}}\ped{r},s)$ and $\mathbf{H}$ as in Eq.~\eqref{eq:Hfunct}, and suppose that $p \in C^1(\mathscr{C};\mathbb{R}_{\geq 0})$ satisfies
\begin{align}\label{eq:condP}
\nabla_{\bm{x}}\cdot p(\bm{x})\bar{\bm{V}}(\bm{x}) \leq 0, \quad \bm{x} \in \mathscr{C}.
\end{align}
Then, the system described by the PDE~\eqref{eq:standard2} (resp.~\eqref{eq:quasiStationary0}) with output~\eqref{eq:Fundef} (resp.~\eqref{eq:Fundef} and~\eqref{eq:Malt}) is passive with storage function
\begin{align}\label{eq:VdissF2}
W\bigl(\bm{u}(\cdot,s)\bigr) \triangleq \dfrac{1}{2}\iint_{\mathscr{C}} p(\bm{x}) \bm{z}^{\mathrm{T}}(\bm{x},s)\bar{\mathbf{K}}_0\bm{z}(\bm{x},s) \dif \bm{x} + \dfrac{1}{2}\sum_{i=1}^n\iint_{\mathscr{C}} p(\bm{x}) \bm{f}^{\mathrm{T}}_i(\bm{x},s)\bar{\mathbf{K}}_i^{-1}\bm{f}_i(\bm{x},s)\dif \bm{x}. 
\end{align}
\begin{proof}
Differentiating Eq.~\eqref{eq:VdissF2} along the dynamics~\eqref{eq:standard2PDE} or~\eqref{eq:quasiStationary0dy} yields
\begin{align}
\begin{split}
\dod{W\bigl(\bm{u}(\cdot,s)\bigr)}{s} & = \iint_{\mathscr{C}}p(\bm{x})\bigggl(\bm{z}^{\mathrm{T}}(\bm{x},s)\bar{\mathbf{K}}_0 + \sum_{i=1}^n \bm{f}_i^{\mathrm{T}}(\bm{x},s)\bigggr)\dpd{\bm{z}(\bm{x},s)}{s} \dif \bm{x} \\
& \quad - \sum_{i=1}^n \iint_{\mathscr{C}}p(\bm{x})\bm{f}_i^{\mathrm{T}}(\bm{x},s)\bar{\mathbf{K}}_i^{-1}\dfrac{\bm{\tau}_i^{-1}}{V\ped{r}}\bm{f}_i(\bm{x},s)\dif \bm{x}  + \sum_{i=1}^n \iint_{\mathscr{C}}p(\bm{x})\bm{f}_i^{\mathrm{T}}(\bm{x},s)\bigl(\bar{\bm{V}}(\bm{x})\cdot \nabla_{\bm{x}}\bigr)\bm{z}(\bm{x},s) \dif \bm{x}\\
& \quad -\dfrac{1}{2}\sum_{i=1}^n\iint_{\mathscr{C}}p(\bm{x})\bar{\bm{V}}(\bm{x})\cdot\nabla_{\bm{x}}\Bigl(\bm{f}_i^{\mathrm{T}}(\bm{x},s)\bar{\mathbf{K}}_i^{-1}\bm{f}_i(\bm{x},s)\Bigr) \dif \bm{x}, \quad s \in (0,S).
\end{split}
\end{align}
Recalling from Eq.~\eqref{eq:ftot} that $\bar{\mathbf{K}}_0\bm{z}(\bm{x},s) + \sum_{i=1}^n \bm{f}_i(\bm{x},s) = \bm{f}(\bm{x},s)$ and using~\eqref{eq:MPsi} gives
\begin{align}\label{eq:WdotSemiFinal}
\begin{split}
\dod{W\bigl(\bm{u}(\cdot,s)\bigr)}{s} & =  -\bigl\langle p(\cdot)\bm{f}(\cdot,s),\bar{\bm{v}}(\cdot,s)\bigr\rangle_{L^2(\mathring{\mathscr{C}};\mathbb{R}^2)} \\
& \quad + \iint_{\mathscr{C}}p(\bm{x})\bm{f}^{\mathrm{T}}(\bm{x},s)\mathbf{\Psi}\bigl(\bar{\bm{v}}(\bm{x},s)\bigr)\bm{f}(\bm{x},s) \dif \bm{x} - \sum_{i=1}^n \iint_{\mathscr{C}}p(\bm{x})\bm{f}_i^{\mathrm{T}}(\bm{x},s)\bar{\mathbf{K}}_i^{-1}\dfrac{\bm{\tau}_i^{-1}}{V\ped{r}}\bm{f}_i(\bm{x},s)\dif \bm{x}\\
& \quad -\dfrac{1}{2}\iint_{\mathscr{C}}p(\bm{x})\bar{\bm{V}}(\bm{x})\cdot\nabla_{\bm{x}}\bigggl( \bm{z}^{\mathrm{T}}(\bm{x},s)\bar{\mathbf{K}}_0\bm{z}(\bm{x},s) + \sum_{i=1}^n\bm{f}_i^{\mathrm{T}}(\bm{x},s)\bar{\mathbf{K}}_i^{-1}\bm{f}_i(\bm{x},s)\bigggr) \dif \bm{x}, \quad s \in (0,S).
\end{split}
\end{align}
Integrating by parts the last term in Eq.~\eqref{eq:WdotSemiFinal} and enforcing the BC~\eqref{eq:BCquasiStandard} or~\eqref{eq:BCquasistesd} leads then to
\begin{align}\label{eq:WdotFinal}
\begin{split}
\dod{W\bigl(\bm{u}(\cdot,s)\bigr)}{s} & \leq  -\bigl\langle p(\cdot)\bm{f}(\cdot,s),\bar{\bm{v}}(\cdot,s)\bigr\rangle_{L^2(\mathring{\mathscr{C}};\mathbb{R}^2)} \\
& \quad + \iint_{\mathscr{C}}p(\bm{x})\bm{f}^{\mathrm{T}}(\bm{x},s)\mathbf{\Psi}\bigl(\bar{\bm{v}}(\bm{x},s)\bigr)\bm{f}(\bm{x},s) \dif \bm{x} - \sum_{i=1}^n \iint_{\mathscr{C}}p(\bm{x})\bm{f}_i^{\mathrm{T}}(\bm{x},s)\bar{\mathbf{K}}_i^{-1}\dfrac{\bm{\tau}_i^{-1}}{V\ped{r}}\bm{f}_i(\bm{x},s)\dif \bm{x}\\
& \quad +\dfrac{1}{2}\iint_{\mathscr{C}}\bigl(\nabla_{\bm{x}} \cdot p(\bm{x})\bar{\bm{V}}(\bm{x})\bigr)\bigggl( \bm{z}^{\mathrm{T}}(\bm{x},s)\bar{\mathbf{K}}_0\bm{z}(\bm{x},s) + \sum_{i=1}^n\bm{f}_i^{\mathrm{T}}(\bm{x},s)\bar{\mathbf{K}}_i^{-1}\bm{f}_i(\bm{x},s)\bigggr) \dif \bm{x}, \quad s \in (0,S).
\end{split}
\end{align}
Since $\bar{\mathbf{K}}_i^{-1}\bm{\tau}_i^{-1} = \bar{\mathbf{C}}_i^{-1} \succeq \mathbf{0}$ for all $i \in \{1,\dots,n\}$, the second and third terms appearing on the right-hand side of Eq.~\eqref{eq:WdotFinal} are negative semidefinite. Therefore, if inequality~\eqref{eq:condP} holds, it may be concluded that 
\begin{align}\label{eq:w03}
-\bigl\langle p(\cdot)\bm{f}(\cdot,s),\bar{\bm{v}}(\cdot,s)\bigr\rangle_{L^2(\mathring{\mathscr{C}};\mathbb{R}^2)} \geq \dod{W\bigl(\bm{u}(\cdot,s)\bigr)}{s}, \quad s \in (0,S).
\end{align}
Finally, integrating the above Eq.~\eqref{eq:w03} proves~\eqref{eq:Fvres} with $w(\bm{f}(\cdot,s),-\bar{\bm{v}}(\cdot,s)) =-\langle p(\cdot)\bm{f}(\cdot,s), \bar{\bm{v}}\ped{r}(\cdot,s)\rangle_{L^2(\mathring{\mathscr{C}};\mathbb{R}^2)}= -\langle p(\cdot)\bm{f}(\cdot,s), \bar{\bm{v}}(\cdot,s)\rangle_{L^2(\mathring{\mathscr{C}};\mathbb{R}^2)}$. 
\end{proof}
\end{lemma}

\begin{lemma}[Passivity of the FrBD$_{n+1}$-GKV models]\label{lemma:DissF202}
Consider the PDE~\eqref{eq:standard2} (resp.~\eqref{eq:quasiStationary0}) with $\mathbf{\Sigma}(\bar{\bm{v}}\ped{r},s)$ and $\mathbf{H}$ as in Eq.~\eqref{eq:Hfunct2}, and suppose that $p \in C^1(\mathscr{C};\mathbb{R}_{\geq 0})$ satisfies~\eqref{eq:condP}.
Then, the system described by the PDE~\eqref{eq:standard2} (resp.~\eqref{eq:quasiStationary0}) with output~\eqref{eq:Fundef} (resp.~\eqref{eq:Fundef} and~\eqref{eq:Malt}) is passive with storage function
\begin{align}\label{eq:VdissF23}
\begin{split}
W\bigl(\bm{u}(\cdot,s)\bigr) & \triangleq \dfrac{1}{2}\iint_{\mathscr{C}} p(\bm{x}) \bigggl(\bm{z}(\bm{x},s)-\sum_{i=1}^n \bm{z}_i(\bm{x},s)\bigggr)^{\mathrm{T}}\bar{\mathbf{K}}_0\bigggl(\bm{z}(\bm{x},s)-\sum_{i=1}^n \bm{z}_i(\bm{x},s)\bigggr) \dif \bm{x} \\
& \quad + \dfrac{1}{2}\sum_{i=1}^n\iint_{\mathscr{C}} p(\bm{x}) \bm{z}^{\mathrm{T}}_i(\bm{x},s)\bar{\mathbf{K}}_i\bm{z}_i(\bm{x},s)\dif \bm{x}. 
\end{split}
\end{align}
\begin{proof}
To prove the result, it is convenient to introduce the auxiliary state $\mathbb{R}^2 \ni \bm{z}_0(\bm{x},s) \triangleq \bm{z}(\bm{x},s)-\sum_{i=1}^n \bm{z}_i(\bm{x},s)$, so that $\bm{f}(\bm{x},s) = \bar{\mathbf{K}}_0\bm{z}_0(\bm{x},s)$. Furthermore, the dynamics of $\bm{z}_0(\bm{x},s)$ is governed by
\begin{subequations}
\begin{align}
\begin{split}
& \dpd{\bm{z}_0(\bm{x},s)}{s} + \bigl(\bar{\bm{V}}(\bm{x},s)\cdot\nabla_{\bm{x}}\bigr)\bm{z}_0(\bm{x},s) =\mathbf{\Psi}\bigl(\bar{\bm{v}}(\bm{x},s)\bigr)\bar{\mathbf{K}}_0\bm{z}_0(\bm{x},s) \\
& \qquad \qquad \qquad \qquad \qquad \qquad \qquad \qquad  - \sum_{i=1}^n \dod{\bm{z}_i(\bm{x},s)}{s} - \bar{\bm{v}}(\bm{x},s), \quad (\bm{x},s) \in \mathring{\mathscr{C}}\times (0,S), 
\end{split} \\
& \bm{z}_0(\bm{x},s) = \bm{0}, \quad (\bm{x},s) \in \mathscr{L}\times (0,S), \\
& \bm{z}_0(\bm{x},0) = \bm{z}_{0,0}(\bm{x}), \quad \bm{x} \in \mathring{\mathscr{C}}. 
\end{align}
\end{subequations}
Accordingly, differentiating Eq.~\eqref{eq:VdissF2} along the dynamics~\eqref{eq:standard2PDE} or~\eqref{eq:quasiStationary0dy} yields
\begin{align}\label{eq:WKVS}
\begin{split}
\dod{W\bigl(\bm{u}(\cdot,s)\bigr)}{s} &= -\bigl\langle p(\cdot)\bm{f}(\cdot,s),\bar{\bm{v}}(\cdot,s)\bigr\rangle_{L^2(\mathring{\mathscr{C}};\mathbb{R}^2)} +\iint_{\mathscr{C}}p(\bm{x})\bm{z}_0^{\mathrm{T}}(\bm{x},s)\bar{\mathbf{K}}_0\mathbf{\Psi}\bigl(\bar{\bm{v}}(\bm{x},s)\bigr)\bar{\mathbf{K}}_0\bm{z}_0(\bm{x},s) \dif \bm{x} \\
& \quad - \sum_{i=1}^n \iint_{\mathscr{C}}p(\bm{x})\bm{z}_0^{\mathrm{T}}(\bm{x},s)\bar{\mathbf{K}}_0\dod{\bm{z}_i(\bm{x},s)}{s} \dif \bm{x} \\
& \quad + \sum_{i=1}^n \iint_{\mathscr{C}} p(\bm{x}) \biggl( \bar{\mathbf{K}}_0\bm{z}_0 - V\ped{r}\bar{\mathbf{C}}_i\dod{\bm{z}_i(\bm{x},s)}{s}\biggr)^{\mathrm{T}}\dod{\bm{z}_i(\bm{x},s)}{s}\dif \bm{x} \\
& \quad  - \dfrac{1}{2}\sum_{i=0}^n \iint_{\mathscr{C}}p(\bm{x})\bar{\bm{V}}(\bm{x})\cdot\nabla_{\bm{x}}\Bigl(\bm{z}_i^{\mathrm{T}}(\bm{x},s)\bar{\mathbf{K}}_i\bm{z}_i(\bm{x},s)\Bigr) \dif \bm{x}, \quad s \in (0,S).
\end{split}
\end{align}
Integrating by parts the last term in Eq.~\eqref{eq:WKVS} and enforcing the BC~\eqref{eq:BCquasiStandard} or~\eqref{eq:BCquasistesd} produces 
\begin{align}\label{eq:WKVS2}
\begin{split}
\dod{W\bigl(\bm{u}(\cdot,s)\bigr)}{s} &\leq -\bigl\langle p(\cdot)\bm{f}(\cdot,s),\bar{\bm{v}}(\cdot,s)\bigr\rangle_{L^2(\mathring{\mathscr{C}};\mathbb{R}^2)} +\iint_{\mathscr{C}}p(\bm{x})\bm{z}_0^{\mathrm{T}}(\bm{x},s)\bar{\mathbf{K}}_0\mathbf{\Psi}\bigl(\bar{\bm{v}}(\bm{x},s)\bigr)\bar{\mathbf{K}}_0\bm{z}_0(\bm{x},s) \dif \bm{x} \\
& \quad - \sum_{i=1}^n \iint_{\mathscr{C}} p(\bm{x})V\ped{r}\dod{\bm{z}_i^{\mathrm{T}}(\bm{x},s)}{s}\bar{\mathbf{C}}_i\dod{\bm{z}_i(\bm{x},s)}{s}\dif \bm{x} \\
& \quad  + \dfrac{1}{2} \sum_{i=0}^n\iint_{\mathscr{C}}\bigl(\nabla_{\bm{x}} \cdot p(\bm{x})\bar{\bm{V}}(\bm{x})\bigr) \bm{z}_i^{\mathrm{T}}(\bm{x},s)\bar{\mathbf{K}}_i\bm{z}_i(\bm{x},s) \dif \bm{x}, \quad s \in (0,S).
\end{split}
\end{align}
The second and third terms appearing on the right-hand side of Eq.~\eqref{eq:WKVS2} are always negative semidefinite. Therefore, if the condition~\eqref{eq:condP} is satisfied, it may be concluded that an inequality formally identical to~\eqref{eq:w03} holds. Hence, the result follows from similar arguments as in the proof of Lemma~\ref{lemma:DissF20}.
\end{proof}
\end{lemma}

Lemmata~\ref{lemma:DissF20} and~\ref{lemma:DissF202} yields passivity for virtually all meaningful parametrisations of~\modref{stdmodel} and~\ref{linmodel2}, concerning both the FrBD$_{n+1}$-GM and FrBD$_{n+1}$-GKV realisations. Some additional remarks are collected below.
\begin{remark}
The inequality in Eq.~\eqref{eq:condP} is the same as that in Lemma 4.2 of \cite{FrBDroll}, and is obviously satisfied when the pressure distribution is constant (for~\modref{stdmodel}, distributions decreasing along the rolling direction also verify Eq.~\eqref{eq:condP} strictly). For viscoelastic materials, contact pressures decreasing along the rolling direction can be explained by considering non-interpenetration constraints. In the literature, spatially varying pressure profiles fulfilling~\eqref{eq:condP} are also employed to reproduce the sign reversal of the steady-state vertical moment at large lateral slips. Essentially, provided that the pressure distribution evolves moderately along the contact area, the rolling process inherits passivity properties from those of the Maxwell and Kelvin-Voigt elements, which are intrinsically dissipative.
\end{remark}

\begin{remark}
If $p(\bm{x}) \geq p\ped{min}$ and $\nabla_{\bm{x}} \cdot p(\bm{x})\bar{\bm{V}}(\bm{x}) \leq -\gamma$, with $p\ped{min}, \gamma \in \mathbb{R}_{>0}$, for all $\bm{x} \in \mathscr{C}$, then the storage functions $W(\bm{u}(\cdot,s))$ in Eqs.~\eqref{eq:VdissF2} and~\eqref{eq:VdissF23} are indeed a Lyapunov functions, and ensure ISS for the PDEs~\eqref{eq:standard2} and~\eqref{eq:quasiStationary0}. In this context, it is worth observing that, for $n=0$, the storage functions~\eqref{eq:VdissF2} and~\eqref{eq:VdissF23} reduce to that obtained for the FrBD$_1$-KV model introduced in \cite{FrBDroll}. In fact, under the condition~\eqref{eq:condP}, passivity properties for the FrBD$_1$-KV model, for which
\begin{align}
\bm{f}(\bm{x},s) = \bar{\mathbf{K}}\bm{z}(\bm{x},s) + V\ped{r}(s)\bar{\mathbf{C}}\dod{\bm{z}(\bm{x},s)}{s},
\end{align}
may be inferred directly from~\eqref{eq:frictionPass} for all $\mathbf{Sym}_2(\mathbb{R}) \ni \bar{\mathbf{K}} \succ \mathbf{0}$, $\mathbf{Sym}_2(\mathbb{R}) \ni \bar{\mathbf{C}} \succeq \mathbf{0}$, and $V\ped{r} \in [V\ped{min},V\ped{max}]$, generalising the findings of \cite{FrBDroll}.
\end{remark}

\begin{remark}
For simplicity, Lemmata~\ref{lemma:DissF20} and~\ref{lemma:DissF202} were proven only for classical solutions ($\varepsilon \in \mathbb{R}_{>0}$ in~\eqref{eq:standard2} and~\eqref{eq:quasiStationary0}); as mentioned in \cite{FrBDroll}, the extension to mild solutions obtained for $\varepsilon \in \mathbb{R}_{\geq0}$ may be established by invoking density and convergence arguments. 
\end{remark}

\section{Numerical simulations}\label{sect:numer}
In the following, FrBD$_2$-GM and FrBD$_3$-GM~\modref{stdmodel},~\ref{semilinmodel}, and~\modref{linmodel2} are numerically simulated and compared to the corresponding FrBD$_1$-KV versions concerning both their steady-state and transient behaviours. More specifically, Sect.~\ref{sect:SS} is dedicated to steady rolling contact, whereas unstationary phenomena are investigated in Sect.~\ref{sect:transient}. As discussed in Appendix~\ref{app:alt}, the FrBD$_2$-GM models correspond directly to a Standard Linear Solid element, for which the interconversion to an equivalent FrBD$_2$-KV parametrisation is available analytically. In the remainder of the paper, the GM formulation is preferred because it is more widely used in the literature, and also because yields a more intuitive state-space reformulation for the FrBD$_2$-SLS dynamics (see Appendix~\ref{app:alt}). Additional implementational and computational details may be found in \cite{FrBDroll}. 

\subsection{Steady-state behaviour}\label{sect:SS}
\begin{figure}
\centering
\includegraphics[width=1\linewidth]{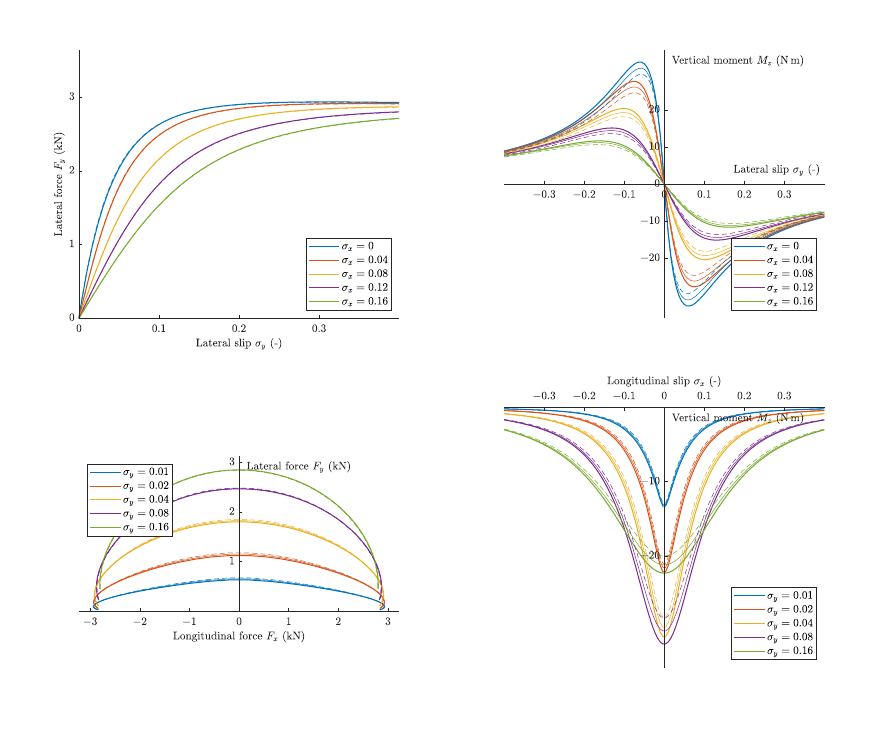} 
\caption{Steady-state characteristics in the absence of spin slips predicted using \modref{semilinmodel} (rectangular contact area with parabolic pressure distribution). Line styles: FrBD$_1$-KV from \cite{FrBDroll} (solid thick lines), FrBD$_2$-GM (solid lines), FrBD$_3$-GM (dashed lines). Model parameters as in Table~\ref{tab:parameters}.}
\label{fig:GoughRect}
\end{figure}

\begin{figure}
\centering
\includegraphics[width=1\linewidth]{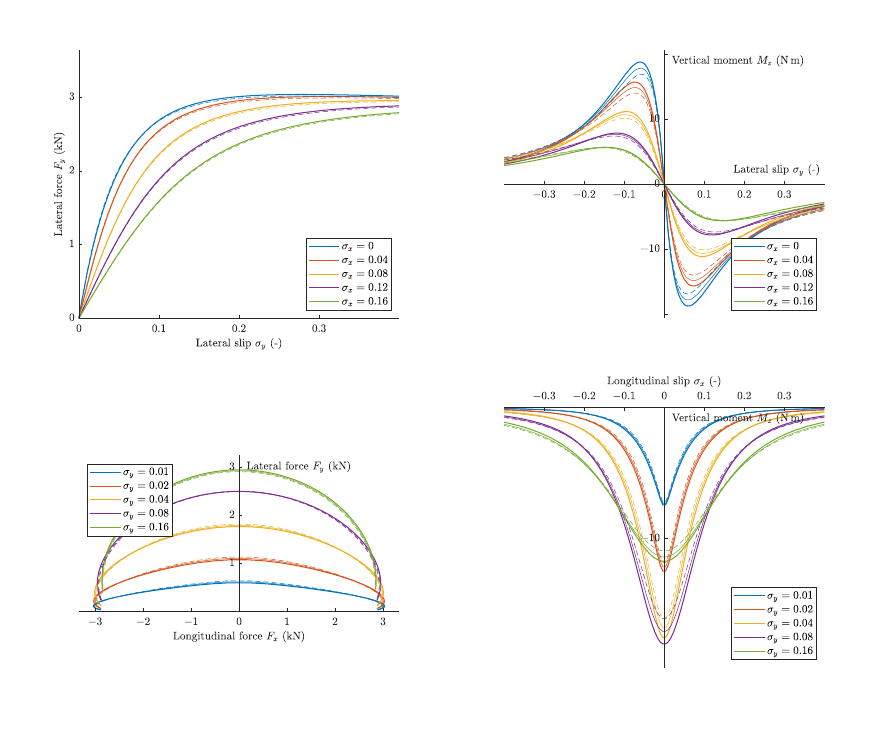} 
\caption{Steady-state characteristics in the absence of spin slips predicted using \modref{semilinmodel} (elliptical contact area with parabolic pressure distribution). Line styles: FrBD$_1$-KV from \cite{FrBDroll} (solid thick lines), FrBD$_2$-GM (solid lines), FrBD$_3$-GM (dashed lines). Model parameters as in Table~\ref{tab:parameters}.}
\label{fig:GoughEll}
\end{figure}

For lumped parameter friction models, all the FrBD$_{n+1}$ variants collapse to the same force-slip relationship given by Eq.~\eqref{eq:fStat}. This confirms analytically that increasing the rheological order does not alter the stationary creep characteristics: the asymptotic friction level is determined solely by the zeroth-order branch stiffness. Physically, this reflects the fact that, under constant sliding, all viscoelastic branches reach equilibrium and no internal stress relaxation remains active.
Due to the distributed nature of the rolling contact process, however, the relaxation phenomena introduced by viscoelastic components in higher-order, distributed parameter rheological representations of the bristle element are not limited to transient regimes, but manifest even under steady-state conditions. Indeed, as suggested already by Eq.~\eqref{eq:condP}, structural damping acts also spatially, contributing to energy dissipation as the system's dynamics evolves along the rolling direction. It is therefore of interest to investigate how more refined viscoelastic descriptions of the bristle dynamics affect the so-called \emph{force-slip surfaces}, namely the mapping $(\hat{\bm{F}}_{\bm{x}},\hat{M}_z) : \mathbb{R}^4 \to \mathbb{R}^3$ given by
\begin{subequations}
\begin{align}
\bm{F}_{\bm{x}} & = \hat{\bm{F}}_{\bm{x}}(\bm{\sigma},\bm{\varphi}), \\
M_z & = \hat{M}_z(\bm{\sigma},\bm{\varphi}).
\end{align}
\end{subequations}
In the absence of spin ($\bm{\varphi} = \bm{0}$), the main characteristics obtained for rectangular and elliptical contact areas with parabolic pressure distributions, and using the model parameters listed in Table~\ref{tab:parameters}, are shown in Figs.~\ref{fig:GoughRect} and~\ref{fig:GoughEll}. The geometrical and rheological parameters reported in Table~\ref{tab:parameters} are of the same order of magnitude as those adopted in previous studies on rubber friction (e.g., \cite{Tsiotras1,Tsiotras3,Deur1,Deur2}) and are consistent with standard values commonly used in the tyre dynamics literature \cite{Guiggiani}.
In particular, the longitudinal and lateral micro-stiffnesses associated with the higher-order branches were selected to be close to those of the zeroth branch. This choice ensures a comparable steady-state response across all model variants, thereby allowing for a clearer assessment of the differences in viscoelastic behaviour arising specifically from the damping and relaxation coefficients. 
Assuming isotropic friction, the matrix $\mathbf{M}(\bm{v}\ped{s})$ was modelled as $\mathbf{M}(\bm{v}\ped{s}) = \mu(\bm{v}\ped{s})\mathbf{I}_2$, with
\begin{align}\label{eq:muExample}
\mu(\bm{v}\ped{s}) = \mu\ped{d} + (\mu\ped{s}-\mu\ped{d})\exp\Biggl(-\biggl(\dfrac{\norm{\bm{v}\ped{s}}_2}{v\ped{S}}\biggr)^{\delta\ped{S}}\Biggr)+ \mu\ped{v}(\bm{v}\ped{s}),
\end{align}
where $\mu\ped{s},\mu\ped{d} \in \mathbb{R}_{>0}$ denote the static and dynamic friction coefficients, $v\ped{S} \in \mathbb{R}_{>0}$ the Stribeck velocity, $\delta\ped{S} \in \mathbb{R}_{\geq 0}$ the Stribeck exponent, and $\mu\ped{v} : \mathbb{R}^2 \to \mathbb{R}_{\geq 0}$ captures the viscous friction. Also in this case, the values of the friction parameters listed in Table~\ref{tab:parameters} were adapted from \cite{Tsiotras1,Tsiotras3,Deur1,Deur2}.

The trends illustrated in Figs.~\ref{fig:GoughRect} and~\ref{fig:GoughEll} are in good qualitative agreement with the force surfaces predicted by other physical-analytical and empirical models of tyre dynamics, such as standard brush models with generalised Coulomb friction and Pacejka's Magic Formula. In particular, adopting a sliding-velocity-dependent friction coefficient as in Eq.~\eqref{eq:muExample} permits reproducing the characteristic force peak at low slip, followed by a progressive decrease in tangential force toward an asymptotic value at large slips.
In this respect, it is worth noting that, whilst LuGre-like formulations of rolling contact friction correctly postulate the friction coefficient as a function of the sliding velocity \cite{Tsiotras1,Tsiotras3,Deur1,Deur2}, Pacejka's Magic Formula and several modified brush models typically assume a slip-dependent equation. As a consequence, the resulting force-slip surfaces are inherently independent of the rolling velocity. However, empirical evidence suggests that the tangential forces and vertical moment do depend on rolling speed \cite{Tsiotras1}, as predicted by sliding-velocity-dependent friction laws similar to Eq.~\eqref{eq:muExample}. Importantly, the capability to consistently combine higher-order viscoelastic rheological behaviour with sliding-velocity-dependent friction is essential in rubber friction applications \cite{Persson1,Persson2,Mavros}. In such cases, slow relaxation dynamics may induce local heating and trigger flash-temperature effects, which in turn modify the frictional response \cite{Persson2,Mavros}. Accounting for these coupled mechanisms is therefore crucial for a physically sound and predictive modelling framework. In this context, it is also worth mentioning that, in Eq.~\eqref{eq:muExample}, additional functional dependencies on the rolling speed may be postulated, for instance by assuming $v\ped{S} = v\ped{S}(V\ped{r})$.

\begin{figure}
\centering
\includegraphics[width=1\linewidth]{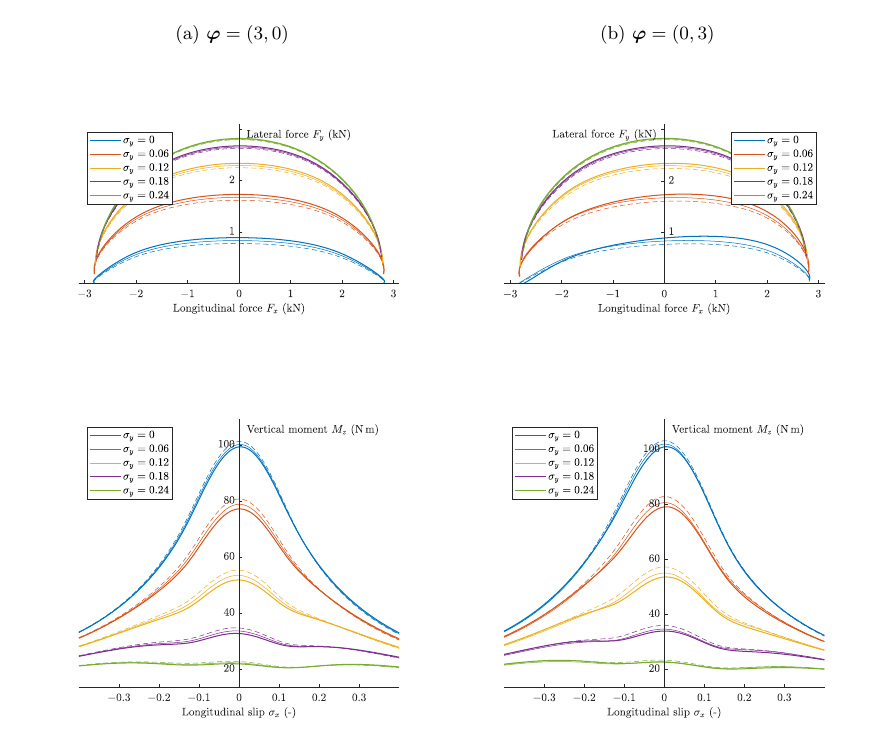} 
\caption{Steady-state characteristics in the presence of large spin slips predicted using \modref{semilinmodel} (rectangular contact area with parabolic pressure distribution): (a) $\bm{\varphi} = (3,0)$; (b) $\bm{\varphi} = (0,3)$. Line styles: FrBD$_1$-KV from \cite{FrBDroll} (solid thick lines), FrBD$_2$-GM (solid lines), FrBD$_3$-GM (dashed lines). Model parameters as in Table~\ref{tab:parameters}.}
\label{fig:RectSpins}
\end{figure}

\begin{figure}
\centering
\includegraphics[width=1\linewidth]{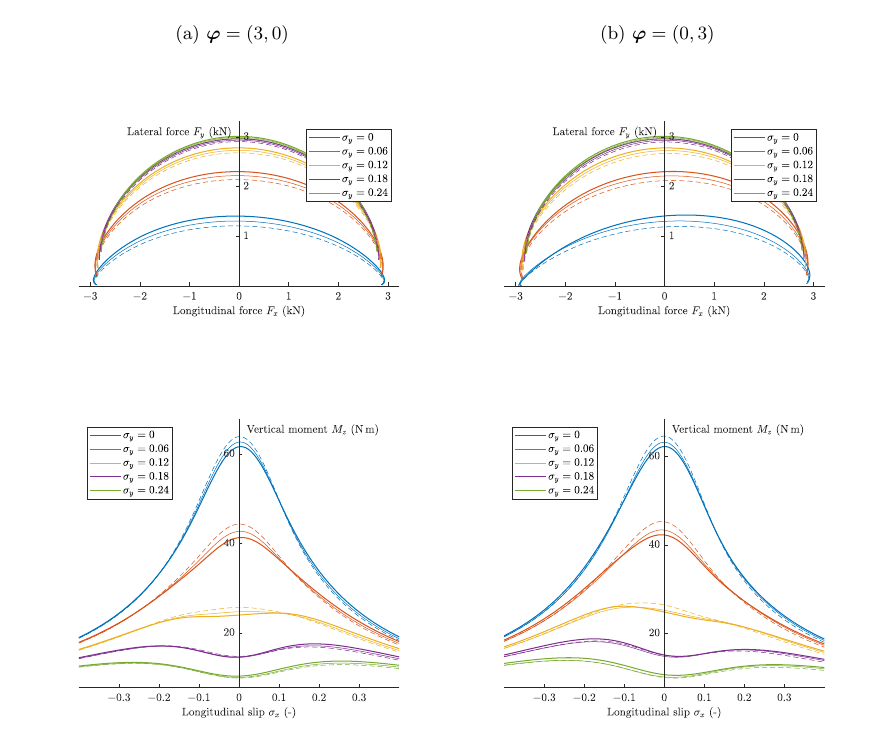} 
\caption{Steady-state characteristics in the presence of large spin slips predicted using \modref{semilinmodel} (elliptical contact area with parabolic pressure distribution): (a) $\bm{\varphi} = (3,0)$; (b) $\bm{\varphi} = (0,3)$. Line styles: FrBD$_1$-KV from \cite{FrBDroll} (solid thick lines), FrBD$_2$-GM (solid lines), FrBD$_3$-GM (dashed lines). Model parameters as in Table~\ref{tab:parameters}.}
\label{fig:EllSpins}
\end{figure}

Local hysteresis affects rolling contact friction also directly: compared with the FrBD$_1$-KV formulation employed in \cite{FrBDroll}, inspection of Figs.~\ref{fig:GoughRect} and~\ref{fig:GoughEll} indicates that the additional relaxation effects introduced by the FrBD$_2$-GM and FrBD$_3$-GM models tend to attenuate both the tangential forces and moments, yielding trends of slightly lower magnitude. Phenomenologically, this may again be explained by observing that, due to the distributed character of the rolling contact process, viscous damping acts spatially along the contact patch length, producing smaller bristle deformations and forces in the proximity of the trailing edge, as also confirmed in simulation. The attenuation is particularly evident in the vertical moment, for both rectangular and elliptical contact areas, where the presence of relaxation over multiple time scales appears to influence not only its magnitude, but also the position of its peak value.
These conclusions remain valid when considering the presence of large spin slips, as illustrated in Figs.~\ref{fig:RectSpins} and~\ref{fig:EllSpins}. In this context, particular attention should be given to the role of viscoelastic relaxation in mitigating the asymmetrisation effects induced by large effective spin values, as observed in both Figs.~\ref{fig:RectSpins}(b) and~\ref{fig:EllSpins}(b). This behaviour is physically intuitive and arises from the same mechanisms discussed above: viscous damping attenuates nonlinear coupling interactions by reducing the bristle deformations, which, in turn, results in a more symmetric trend of the lateral and longitudinal forces. 

Additional numerical investigations, not reported here for the sake of brevity, were carried out by varying the rheological parameters listed in Table~\ref{tab:parameters} whilst keeping $n=1$ and $n=2$ fixed. These parametric studies did not reveal any substantial differences in the predicted response, indicating that the model behaviour is essentially determined by the rheological order, but relatively insensitive to moderate variations of the adopted coefficients.

\begin{table}[h!]\centering 
\caption{Model parameters. Matrices $\bar{\mathbf{K}}_0$, $\bar{\mathbf{K}}_1$, $\bar{\mathbf{K}}_2$, $\bm{\tau}_1$, and $\bm{\tau}_2$ as in Eq.~\eqref{eq:Sigmas2}; $\mathbf{M}(\bm{y}) = \mu(\bm{y})\mathbf{I}_2$ with $\mu(\bm{y})$ as in Eq.~\eqref{eq:muExample}}
{\begin{tabular}{|c|l|c|c|}
\hline
Parameter & Description & Unit & Value \\
\hline 
$a$ & Contact area semilength & m & 0.075\\
$b$ & Contact area semiwidth & m & 0.05\\
$\bar{k}_{0x}$ &Longitudinal micro-stiffness (0th branch) & $\textnormal{m}^{-1}$ & 240 \\
$\bar{k}_{0y}$ & Lateral micro-stiffness (0th branch) & $\textnormal{m}^{-1}$ & 240 \\
$\bar{k}_{1x}$ &Longitudinal micro-stiffness (1st branch) & $\textnormal{m}^{-1}$ & 240 \\
$\bar{k}_{1y}$ & Lateral micro-stiffness (1st branch) & $\textnormal{m}^{-1}$ & 240 \\
$\bar{k}_{2x}$ &Longitudinal micro-stiffness (2nd branch) & $\textnormal{m}^{-1}$ & 260 \\
$\bar{k}_{2y}$ & Lateral micro-stiffness (2nd branch) & $\textnormal{m}^{-1}$ & 260 \\
$\tau_{1x}$ &Longitudinal relaxation time (1st branch) & s &0.1 \\
$\tau_{1y}$ & Lateral relaxation time (1st branch) & s & 0.1 \\
$\tau_{2x}$ &Longitudinal relaxation time (2nd branch)  & s &0.05 \\
$\tau_{2y}$ & Lateral relaxation time (2nd branch) & s & 0.05 \\
$\mu\ped{s}$ & Static friction coefficient & -& 1\\
$\mu\ped{d}$ & Dynamic friction coefficient & -& 0.7\\
$\mu\ped{v}$ & Viscous friction coefficient & -& 0\\
$v\ped{S}$ & Stribeck velocity & $\textnormal{m}\,\textnormal{s}^{-1}$ & 3.49 \\
$\delta\ped{S}$ & Stribeck exponent & - & 0.6 \\
$F_{z}$ & Vertical force & N & 4000 \\
$\varepsilon$ & Regularisation parameter & $\textnormal{m}^2\,\textnormal{s}^{-2}$ & $10^{-12}$ \\ 
\hline
\end{tabular} }
\label{tab:parameters}
\end{table}
\subsection{Transient behaviour}\label{sect:transient}
Relaxation phenomena play a critical role in rolling contact. In tyre dynamics, for instance, relaxation effects associated with the local evolution of stresses and deformations within the contact patch can trigger unstable behaviours in road vehicles, as confirmed both theoretically and experimentally in a robust body of literature \cite{Takacs1,Takacs2,Takacs3,Takacs4,Takacs5,BicyclePDE,SemilinearV}.

In tyres, relaxation arises from three main sources: (i) flexibility of secondary structural components, such as the carcass and sidewall \cite{Higuchi2,Rill1,PAC}; (ii) microscopic transient phenomena, which produce dynamics akin to finite-time delays \cite{USB,Meccanica2,SphericalWheel,FrBDroll}; and (iii) polymeric dissipation, effectively captured by higher-order rheological elements such as those considered in this manuscript. The following qualitative analysis focuses on the latter, examining how refined descriptions of the bristle dynamics affect the unsteady behaviour of rolling contact forces and moments. For simplicity, the discussion is restricted to translational slip inputs, as spin slips have only marginal influence upon the general transient dynamics \cite{FrBDroll}.

\begin{figure}
\centering
\includegraphics[width=0.9\linewidth]{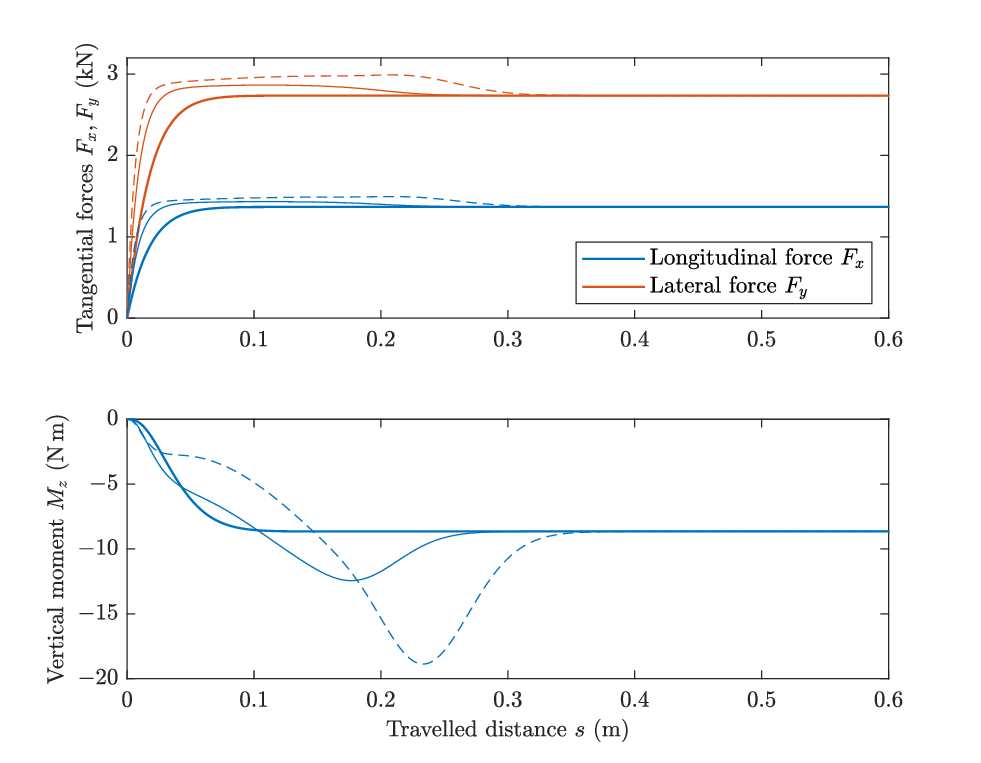} 
\caption{Transient characteristics predicted by~\modref{stdmodel} for a step slip input $\bm{\sigma} = (0.08, 0.16)$ in the absence of spin (line contact with parabolic pressure distribution). Line styles: FrBD$_1$-KV from \cite{FrBDroll} (solid thick lines), FrBD$_2$-GM (solid lines), FrBD$_3$-GM (dashed lines). Model parameters as in Table~\ref{tab:parameters}.}
\label{fig:trans1}
\end{figure}

First, the system's response to a step input in the rigid relative velocity is considered. Figure~\ref{fig:trans1}, produced using \modref{stdmodel} in line contact, illustrates the tangential forces and vertical moment for a step slip input $\bm{\sigma} = (0.08, 0.16)$. The FrBD$_1$-KV model (solid thick lines) exhibits a response closely resembling a first-order system in all characteristics. In contrast, the higher-order FrBD$_2$-GM and FrBD$_3$-GM models (solid and dashed lines) display richer dynamics, including an initial overshoot characteristic of nonlinear systems. Notably, whilst the FrBD$_1$-KV formulation predicts convergence of the forces and moment to their steady-state values for $s \leq 0.1$ m, the refined models require approximately $s = 0.3$ m. This highlights the significant role of hysteretic damping in shaping the transient response of rolling contact elements, including pneumatic and non-pneumatic tyres. An interpretation of the behaviour depicted in Fig.~\ref{fig:trans1} may be obtained by observing that, in the distributed formulation, time relaxation is convected spatially along the contact patch. For a constant rolling speed $V\ped{r}(s) = V\ped{r}$, each relaxation time $\bm{\tau}_i$, $i \in \{1,\dots,n\}$, corresponds to a characteristic relaxation length. When the imposed slip variation occurs over distances comparable to or smaller than these relaxation lengths, the dissipative branches cannot fully equilibrate before exiting the contact region. As a consequence, elastic energy is temporarily stored in the slower branches and released downstream, producing the observed overshoot and delayed convergence. The FrBD$_1$-KV model, being associated with a single relaxation time, can reproduce only one characteristic length scale and therefore cannot capture the multi-stage adjustment process emerging in the FrBD$_2$ and FrBD$_3$ formulations.

\begin{figure}
\centering
\includegraphics[width=0.9\linewidth]{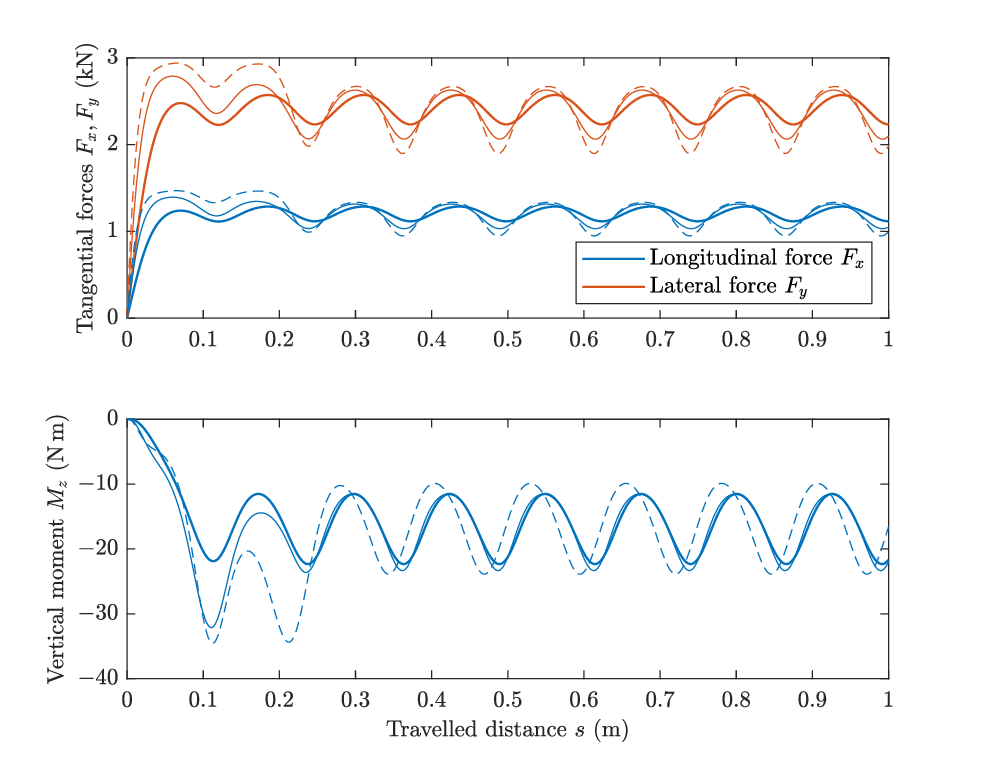} 
\caption{Transient forces predicted by~\modref{linmodel2} for a sinusoidal slip input as in Eq.~\eqref{eq:sigmaSIn}, with $\bm{\sigma} = (0.04,0.08)$ and $(\omega_1,\omega_2) = (50,0)$ $\mathrm{m}^{-1}$ (line contact with parabolic pressure distribution). Line styles: FrBD$_1$-KV from \cite{FrBDroll} (solid thick lines), FrBD$_2$-GM (solid lines), FrBD$_3$-GM (dashed lines). Model parameters as in Table~\ref{tab:parameters}.}
\label{fig:trans2}
\end{figure}

\begin{figure}
\centering
\includegraphics[width=0.9\linewidth]{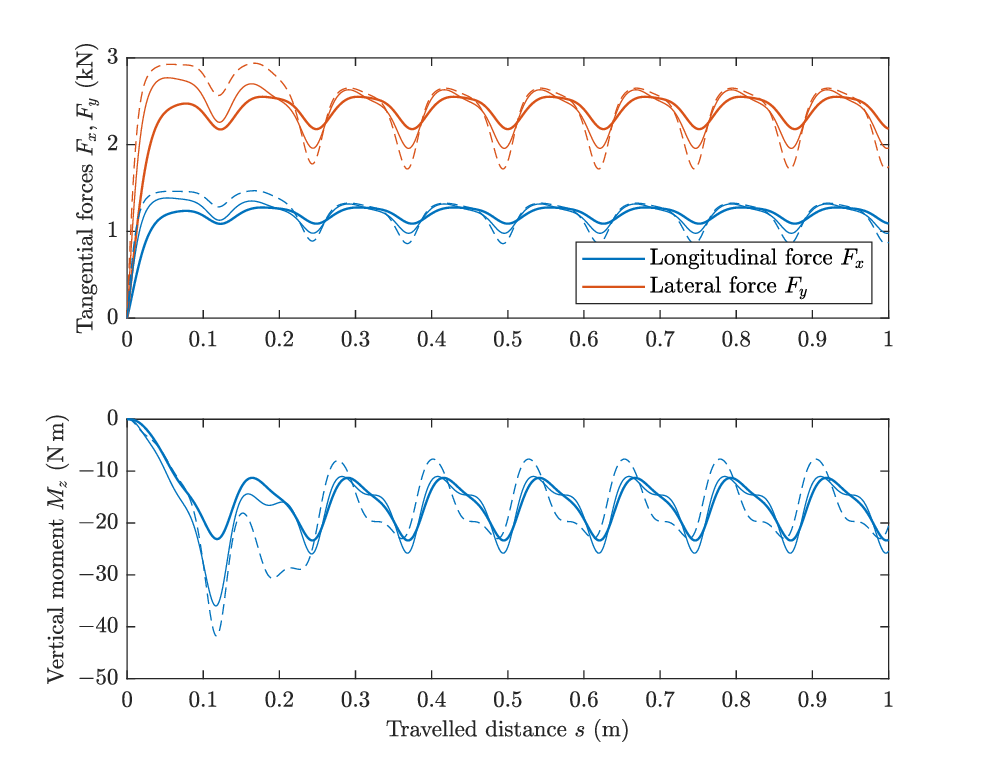} 
\caption{Transient forces predicted by~\modref{linmodel2} for a sinusoidal slip input as in Eq.~\eqref{eq:sigmaSIn}, with $\bm{\sigma} = (0.04,0.08)$ and $(\omega_1,\omega_2) = (50,100)$ $\mathrm{m}^{-1}$ (line contact with parabolic pressure distribution). Line styles: FrBD$_1$-KV from \cite{FrBDroll} (solid thick lines), FrBD$_2$-GM (solid lines), FrBD$_3$-GM (dashed lines). Model parameters as in Table~\ref{tab:parameters}.}
\label{fig:trans3}
\end{figure}

Additional simulations were conducted to investigate the rolling contact system's response to a sinusoidal slip input, generally modelled as
\begin{align}\label{eq:sigmaSIn}
\bm{\sigma}(s) = \bar{\bm{\sigma}}\bigl[1 + 0.5\sin(\omega_1 s) + 0.25\sin(\omega_2 s)\bigr], \quad s\in [0,S].
\end{align}
Figures~\ref{fig:trans2} and~\ref{fig:trans3} visualise the tangential forces and vertical moment for $\bar{\bm{\sigma}} = (0.04, 0.08)$ and $(\omega_1,\omega_2) = (50,0)$, $(\omega_1,\omega_2) = (50,100)$ $\textnormal{m}^{-1}$, respectively. The trends confirm the previous analysis: whilst the simplest model reproduces the input oscillations almost immediately, the more sophisticated FrBD$_2$-GM and FrBD$_3$-GM introduce additional delay effects that are not adequately captured by standard rheological representations. This discrepancy is particularly pronounced for the vertical moment, which is consistent with the observations reported in Sect.~\ref{sect:SS}. From a frequency-domain viewpoint, the observed differences may be explained by recalling that a GM element exhibits a storage and loss modulus that vary with excitation frequency (see Appendix~\ref{app:param}). The sinusoidal slip input, therefore, probes the effective complex modulus of the bristle element. Whilst the FrBD$_1$-KV model behaves essentially as a single-pole low-pass filter, the higher-order GM representations introduce multiple poles associated with distinct relaxation times, resulting in amplitude attenuation and phase lag that depend on the spectral content of the imposed slip variation.

The phenomena discussed above have profound repercussions on the dynamics of road vehicles. Indeed, they imply that, in tyres, rapid manoeuvres or high-frequency excitation -- such as those encountered during braking on rough surfaces or wheel shimmy phenomena -- may activate viscoelastic mechanisms that remain essentially invisible in first-order models. More generally, the rheological order becomes particularly relevant when the characteristic time scales of vehicle manoeuvres are comparable to the intrinsic material relaxation times. Under slow quasi-static conditions, all formulations collapse to nearly identical responses; however, in highly dynamic scenarios, the simplified FrBD$_1$-KV model may underestimate transient peaks in forces and moments. Since such peaks are known to influence stability margins in lateral dynamics and braking performance, incorporating experimentally identified relaxation spectra may improve predictive accuracy without altering the underlying friction law. 
 
\section{Conclusions}\label{sect:conclusion}
The present paper significantly extended the recently developed rFrBD framework, providing a unified and systematic treatment of viscoelastic rolling contact. This was achieved by integrating the two most advanced rheological descriptions of viscoelastic solids available within the classical derivatives framework: the Generalised Maxwell (GM) and Generalised Kelvin-Voigt (GKV) models. By combining standard elements in series or parallel, these representations can accurately capture complex multi-frequency relaxation behaviours that simpler models are unable to reproduce.

The resulting FrBD models, renamed FrBD$_{n+1}$, were formulated as systems of $2(n+1)$ hyperbolic PDEs, describing the distributed evolution of total bristle deflection, friction force, and internal stress and deformation variables. This formulation enables realistic representation of spatially varying phenomena in the contact region, incorporating local kinematics, transport effects, and spin. 
Following \cite{FrBDroll}, three model variants of increasing complexity, accounting for different levels of spin excitation, were derived from both the GM- and GKV-based formulations. The linear versions were analysed with respect to key mathematical properties, including well-posedness and passivity. Notably, it was shown that the transition from linear differential rheological models to the corresponding frictional PDEs preserves passivity for virtually any physically meaningful parameterisation. This not only guarantees the passivity of the rolling contact process but also corroborates the theoretical soundness of the developed friction models, which are always consistent with the physical reality.

The theoretical developments of this work were complemented by numerical simulations that illustrated the influence of higher-order rheological descriptions and their associated relaxation effects on both steady-state and transient behaviours. In particular, it was qualitatively demonstrated that the hysteretic phenomena introduced by dissipative components correctly accounted for by GM and GKV models can significantly affect rolling dynamics even in steady state, since the structural damping contributions act across both time and space. The trend of the generated vertical moment appears to be especially influenced by these internal relaxation effects, demanding the adoption of refined viscoelastic descriptions. Moreover, relaxation processes distributed over multiple time scales seem to excite dynamical responses that are typical of nonlinear systems. Higher-order viscoelastic dynamics may thus be expected to influence the stability and transient behaviour of interconnected mechanical systems -- including road and railway vehicles -- where rolling contact is a key interaction mechanism.

Compared to classical hereditary viscoelastic rolling formulations, which involve time convolution integrals, the approach presented in this paper provides a state-space realisation with explicit internal variables, facilitating coupling with multibody simulations and control algorithms while preserving consistency with linear viscoelastic material theory. In particular, in contrast to phenomenological tyre brush models, where relaxation effects are embedded in empirical relaxation lengths, the FrBD$_{n+1}$ framework links transient rolling behaviour directly to experimentally identifiable material spectra.
Finally, it should be emphasised that the models developed in this manuscript are exclusively based on classical derivative rheological descriptions. Recent studies have shown that fractional-derivative models can successfully capture viscoelasticity with fewer parameters. Future work may therefore focus on deriving more sophisticated formulations that combine PDEs with fractional ODEs, providing physically and thermodynamically consistent descriptions of viscoelastic rolling contact.


\section*{Funding declaration}
This research was financially supported by the project FASTEST (Reg. no. 2023-06511), funded by the Swedish Research Council. 


\section*{Compliance with Ethical Standards}

The authors declare that they have no conflict of interest.

\section*{Author Contribution declaration}
L.R. is the sole author and contributor to the manuscript.

\appendix
\section{Model parametrisation}\label{app:param}

This appendix outlines a practical procedure for parametrising the FrBD$_{n+1}$ models based on
Generalised Maxwell (GM) and Generalised Kelvin-Voigt (GKV) rheological representations from
frequency-domain experimental data. The discussion is restricted to linear viscoelasticity and aims
at bridging the theoretical formulation presented in Sect.~\ref{sect:2Dext} with commonly employed experimental
characterisation techniques.

\subsection{Linear viscoelastic response in the frequency domain}
In the present formulation, the bristle deformation $\bm{z}(t) \in \mathbb{R}^2$ and the corresponding force
$p\bm{f}(t) \in \mathbb{R}^2$ are vector-valued quantities. Accordingly, the linear viscoelastic response is
described by a matrix-valued complex modulus $\mathbf{G}^*(\omega) \in \mathbf{Sym}_2(\mathbb{C})$. For a harmonic deformation $\bm{z}(t) = \RE\{ \hat{\bm{z}} \eu^{\im\omega t} \}$, the resulting force response reads
$p\bm{f}(t) = \RE\{ \mathbf{G}^*(\omega)\hat{\bm{z}} e^{\im\omega t} \}$. The real and imaginary parts of $\mathbf{G}^*(\omega)$,
\begin{align}
\mathbf{G}^\prime(\omega) \triangleq \RE\bigl\{\mathbf{G}^*(\omega)\bigr\}, \quad \textnormal{and} \quad \mathbf{G}^{\prime \prime}(\omega) \triangleq \IM\bigl\{\mathbf{G}^*(\omega)\bigr\},
\end{align}
are referred to as the storage and loss moduli \cite{Rheology1,bookRheol1}, respectively, and belong to
$\mathbf{Sym}_2(\mathbb{R})$.
In many practical applications, experimental characterisation is performed independently in the longitudinal and lateral directions. In such cases, $\mathbf{G}^*(\omega)$ is taken diagonal, and the storage and loss moduli reduce to scalar functions along each principal direction. The present formulation, however, also accommodates coupled or anisotropic viscoelastic responses by allowing for full symmetric matrices.
For a GM element with $n+1$ branches, the complex modulus admits the representation
\begin{align}
\mathbf{G}^*(\omega) =\mathbf{K}_0 + \sum_{i=1}^{n}\im \omega \mathbf{C}_i(\mathbf{K}_i + \im\omega\mathbf{C}_i)^{-1}\mathbf{K}_i,
\label{eq:GM_complex_modulus}
\end{align}
where $\mathbf{Sym}_2(\mathbb{R}) \ni \mathbf{K}_i \succ \mathbf{0}$, $i \in \{0,\dots,n\}$, and $\mathbf{Sym}_2(\mathbb{R}) \ni \mathbf{C}_i \succ \mathbf{0}$, $i \in \{1,\dots,n\}$, denote stiffness and damping matrices, i.e., $\mathbf{K}_i = p\bar{\mathbf{K}}_i$, and $\mathbf{C}_i = p\bar{\mathbf{C}}_i$.
An equivalent expression can be derived for the GKV representation, yielding the same complex modulus
$\mathbf{G}^*(\omega)$, in accordance with the equivalence stated in Eq.~\eqref{eq:rheol1_diff}. Equation~\eqref{eq:GM_complex_modulus} highlights that GM and GKV elements naturally generate
a discrete relaxation spectrum, with characteristic relaxation time matrices
$\bm{\tau}_i = \bar{\mathbf{C}}_i \bar{\mathbf{K}}_i^{-1} = \mathbf{C}_i \mathbf{K}_i^{-1}$, $i \in \{1,\dots,n\}$.

\subsection{Identification from frequency-domain measurements}

In practice, the viscoelastic properties of polymers and rubbers are commonly identified using dynamic mechanical analysis (DMA) \cite{Rheology1,Rheology2}, which provides measurements of $\mathbf{G}^\prime(\omega)$ and $\mathbf{G}^{\prime \prime}(\omega)$ over a prescribed frequency range. The identification of a GM or GKV model then amounts to fitting
Eq.~\eqref{eq:GM_complex_modulus} to the experimentally measured complex modulus.

A standard identification pipeline consists of: (i) selecting \emph{a priori} a finite number $n$ of relaxation branches; (ii) assuming $\bm{\tau}_i = \tau_i\mathbf{I}_2$, $i \in \{1,\dots,n\}$, prescribing a set of relaxation times $\{\tau_i\}_{i=1}^n$; (iii) estimating the matrices $\mathbf{K}_i$ and $\mathbf{C}_i$ via least-squares fitting of $\mathbf{G}^*(\omega)$. In most engineering applications, a small number of branches (typically $n = 2$-$5$) suffices to accurately reproduce the viscoelastic response over the frequency range relevant to rolling contact phenomena.
In particular, the relaxation times $\tau_i$ determine the frequency range over which viscoelastic dissipation
occurs. A common and effective strategy consists in distributing them logarithmically over a
prescribed interval $[\tau\ped{min}, \tau\ped{max}]$, with $\tau\ped{min}, \tau\ped{max} \in \mathbb{R}_{>0}$, according to
\begin{align}
\tau_i = \tau\ped{min} \biggl(\frac{\tau\ped{max}}{\tau\ped{min}}\biggr)^{\frac{i-1}{n-1}}, \quad i \in \{1,\dots,n\}.
\label{eq:log_spaced_taus}
\end{align}
Short relaxation times capture high-frequency, rate-dependent effects, whereas long relaxation times govern slow stress relaxation and hysteresis. The bounds $\tau\ped{min}$ and $\tau\ped{max}$ may be chosen based on the expected range of excitation frequencies induced by rolling, which depend on rolling speed, contact patch dimensions, and slip conditions.

\subsection{FrBD$_{n+1}$ model parameters and interconversions}

Once the stiffness and damping matrices $\mathbf{K}_i$ and $\mathbf{C}_i$ have been identified, they can be directly employed in the FrBD$_{n+1}$ models introduced in Sect.~\ref{sect:DynamicDer} by introducing the normalised parameters $\bar{\mathbf{K}}_i = \mathbf{K}_i/p$, $i \in \{0,\dots,n\}$, and $\bar{\mathbf{C}}_i = \mathbf{C}_i/p$, $i \in \{1,\dots,n\}$. where $p(\bm{x},s)$ denotes the local normal pressure.
For the GM realisation, these parameters enter the dynamics through Eq.~\eqref{eq:ODEModel}, whereas for the GKV realisation they appear in Eq.~\eqref{eq:ODEModelKV}. 

In the developments of Sect.~\ref{sect:2Dext}, the normalised matrices $\bar{\mathbf{K}}_i$ and $\bar{\mathbf{C}}_i$ have been assumed to be spatially constant. This assumption is automatically satisfied in the case of uniform pressure distributions. In practical rolling-contact applications, however, mildly non-uniform pressure distributions, typically decreasing along the rolling direction, are often preferred, as they permit reproducing the sign reversal of the vertical moment at large slips \cite{Tsiotras3,Deur1,Deur2,FrBDroll}, and ensure strict passivity of the distributed models (Eqs.~\eqref{eq:VdissF2} and~\eqref{eq:VdissF23} become Lyapunov functions). In such cases, it is convenient to allow the original matrices $\mathbf{K}_i$ and $\mathbf{C}_i$ to vary proportionally with the local pressure, so that the corresponding normalised matrices remain constant over the contact patch.

Finally, additional comments are provided regarding possible model interconversions.
For any finite $n$, the GM and GKV representations are equivalent, in the sense that they generate the
same constitutive relation between force and deformation \cite{Rheology2,Rheology2-3}, as expressed in Eq.~\eqref{eq:rheol1_diff} (provided that the matrices $\bar{\mathbf{K}}_i$ and $\bar{\mathbf{C}}_i$ are constant). Whilst closed-form interconversion formulas become increasingly involved as $n$ grows, both representations yield
identical input-output behaviour. However, from a numerical perspective, the GM realisation introduces internal force states $\bm{f}_i(\bm{x},s)$, whereas the GKV formulation employs internal deformation states $\bm{z}_i(\bm{x},s)$. The choice between the two may therefore be guided by considerations related to numerical conditioning or implementation convenience. In this context, it should be mentioned that fractional viscoelastic models may also be interpreted as the limit of GM or GKV models with a continuous relaxation spectrum \cite{Rheology3}. In this sense, the FrBD$_{n+1}$ framework provides a finite-dimensional approximation of fractional constitutive laws, whilst retaining a clear physical interpretation and a state-space structure amenable to analysis. Increasing the number of dissipative branches $n$ allows approximating increasingly complex relaxation behaviours without resorting to fractional derivatives.

\section{Alternative state-space representations of the FrBD$_2$-SLS model}\label{app:alt}
The present appendix discusses some alternative representation of the FrBD$_2$-SLS model, which is based on a Standard Linear Solid rheological bristle element, and may be obtained from certain parametrisations of Eqs.~\eqref{eq:ODEModel} and~\eqref{eq:ODEModelKV} for $n = 1$.

Starting with the FrBD$_2$-GM, Eq.~\eqref{eq:ODEModel} for $n = 1$, admits an alternative state-space representation as
\begin{subequations}\label{eq:ODEModelSLS}
\begin{align}
& \dot{\bm{z}}(t) = -\mathbf{M}^{-2}\bigl(\bm{v}\ped{r}(t)\bigr)\norm{\mathbf{M}\bigl(\bm{v}\ped{r}(t)\bigr)\bm{v}\ped{r}(t)}_{2,\varepsilon}\bm{f}(t)-\bm{v}\ped{r}(t), \\
& \dot{\bm{f}}(t) = -\bm{\tau}^{-1}\bm{f}(t) + \bm{\tau}^{-1}\bar{\mathbf{K}}_0\bm{z}(t)+ \bigl(\bar{\mathbf{K}}_0 +\bar{\mathbf{K}}_1\bigr)\dot{\bm{z}}(t), \quad t \in (0,T),\\
& \bm{z}(0) = \bm{z}_0, \; \bm{f}(0) = \bm{f}_{0},
\end{align}
\end{subequations}
where, for notational convenience, it has been renamed $\mathbf{Sym}_2(\mathbb{R}) \ni \bm{\tau} \triangleq \bm{\tau}_1 = \bar{\mathbf{C}}_1\bar{\mathbf{K}}_1^{-1}$. From the above Eq.~\eqref{eq:ODEModelSLS}, it may be easily realised that, for $\bar{\mathbf{C}}_1 = \mathbf{0}$, the FrBD$_2$-SLS model degenerates into the Dahl one, whereas the FrBD$_1$-KV formulation introduced in \cite{FrBDroll}, corresponding to an amended LuGre, is recovered when $\norm{\bar{\mathbf{K}}_1} \gg \norm{\bar{\mathbf{K}}_0}$.

Defining the distributed state as $\mathbb{R}^4 \ni \bm{u}(\bm{x},s)\triangleq [\bm{z}^{\mathrm{T}}(\bm{x},s) \; \bm{f}^{\mathrm{T}}(\bm{x},s)]^{\mathrm{T}}$, the distributed rolling contact model may be put in the form~\eqref{eq:PDenoMOdel} with
\begin{subequations}
\begin{align}
\mathbf{\Sigma}(\bar{\bm{v}}\ped{r},s) & \triangleq \begin{bmatrix} \mathbf{0} & \mathbf{\Psi}(\bar{\bm{v}}\ped{r},s) \\
\dfrac{\bm{\tau}^{-1}}{V\ped{r}(s)}\bar{\mathbf{K}}_0 &-\dfrac{\bm{\tau}^{-1}}{V\ped{r}(s)}+ (\bar{\mathbf{K}}_0 + \bar{\mathbf{K}}_1)\mathbf{\Psi}(\bar{\bm{v}}\ped{r},s)\end{bmatrix}, \\
\mathbf{H} & \triangleq -\begin{bmatrix} \mathbf{I}_2 & \bar{\mathbf{K}}_0 + \bar{\mathbf{K}}_1 \end{bmatrix}^{\mathrm{T}}.
\end{align}
\end{subequations}
The corresponding storage function in Eq.~\eqref{eq:VdissF2} rewrites accordingly
\begin{align}\label{eq:VdissF2SLS}
\begin{split}
W\bigl(\bm{u}(\cdot,s)\bigr) & \triangleq \dfrac{1}{2}\iint_{\mathscr{C}} p(\bm{x}) \bm{z}^{\mathrm{T}}(\bm{x},s)\bar{\mathbf{K}}_0\bm{z}(\bm{x},s) \dif \bm{x} \\
& \quad + \dfrac{1}{2}\iint_{\mathscr{C}} p(\bm{x}) \bigl(\bm{f}(\bm{x},s)-\bar{\mathbf{K}}_0\bm{z}(\bm{x},s)\bigr)^{\mathrm{T}}\bar{\mathbf{K}}_1^{-1}\bigl(\bm{f}(\bm{x},s)-\bar{\mathbf{K}}_0\bm{z}(\bm{x},s)\bigr)\dif \bm{x}. 
\end{split}
\end{align}
On the other hand, introducing the variable $\mathbb{R}\ni\bm{z}_0(t) \triangleq \bm{z}(t)-\bm{z}_1(t)$, Eq.~\eqref{eq:ODEModelKV} can be recast as
\begin{subequations}\label{eq:ODEModelSLS2}
\begin{align}
& \dot{\bm{z}}_0(t) = -\mathbf{M}^{-2}\bigl(\bm{v}\ped{r}(t)\bigr)\norm{\mathbf{M}\bigl(\bm{v}\ped{r}(t)\bigr)\bm{v}\ped{r}(t)}_{2,\varepsilon}\bar{\mathbf{K}}_0\bm{z}_0-\dot{\bm{z}}_1(t)-\bm{v}\ped{r}(t), \\
& \dot{\bm{z}}_1(t) = -\bar{\mathbf{C}}_1^{-1}\bar{\mathbf{K}}_1\bm{z}_1(t) + \bar{\mathbf{C}}_1^{-1}\bar{\mathbf{K}}_0\bm{z}_0(t), \quad t \in (0,T),\\
& \bm{z}_0(0) = \bm{z}_{0,0}, \; \bm{z}_1(0) = \bm{z}_{1,0},
\end{align}
\end{subequations}
With the distributed state vector $\mathbb{R}^4\ni \bm{u}(\bm{x},s) \triangleq [\bm{z}_0^{\mathrm{T}}(\bm{x},s)\; \bm{z}_1^{\mathrm{T}}(\bm{x},s)]^{\mathrm{T}}$, the distributed rolling contact model may be put in the form~\eqref{eq:PDenoMOdel} with
\begin{subequations}
\begin{align}
\mathbf{\Sigma}(\bar{\bm{v}}\ped{r},s) & \triangleq \begin{bmatrix} \Biggl(\mathbf{\Psi}(\bar{\bm{v}}\ped{r},s)-\dfrac{\bar{\mathbf{C}}_1^{-1}}{V\ped{r}(s)}\Biggr)\bar{\mathbf{K}}_0 & \dfrac{\bar{\mathbf{C}}_1^{-1}}{V\ped{r}(s)}\bar{\mathbf{K}}_1 \\
\dfrac{\bar{\mathbf{C}}_1^{-1}}{V\ped{r}(s)}\bar{\mathbf{K}}_0 & -\dfrac{\bar{\mathbf{C}}_1^{-1}}{V\ped{r}(s)}\bar{\mathbf{K}}_1 \end{bmatrix}, \\
\mathbf{H} & \triangleq -\begin{bmatrix} \mathbf{I}_2 & \mathbf{0} \end{bmatrix}^{\mathrm{T}}.
\end{align}
\end{subequations}
The corresponding storage function in Eq.~\eqref{eq:VdissF23} rewrites accordingly
\begin{align}\label{eq:VdissF2SLS3}
\begin{split}
W\bigl(\bm{u}(\cdot,s)\bigr) & \triangleq \dfrac{1}{2}\iint_{\mathscr{C}} p(\bm{x}) \bm{z}_0^{\mathrm{T}}(\bm{x},s)\bar{\mathbf{K}}_0\bm{z}_0(\bm{x},s) \dif \bm{x} +\dfrac{1}{2}\iint_{\mathscr{C}} p(\bm{x}) \bm{z}_1^{\mathrm{T}}(\bm{x},s)\bar{\mathbf{K}}_1\bm{z}_1(\bm{x},s) \dif \bm{x}. 
\end{split}
\end{align}

\newpage
\textbf{Errata} \newline

Errata to "Romano, L. Two-dimensional FrBD friction models for rolling contact: Extension to linear viscoelasticity. Tribology International 220, 111953 (2026). https://doi.org/10.1016/j.triboint.2026.111953". 



\begin{enumerate}
\item The statement of Theorem 4.1 was modified as follows:

Suppose that $\mathring{\mathscr{C}}\subset \mathbb{R}^2$ is bounded, with boundary $\partial \mathscr{C}$ piecewise $C^1$. Then, for all $\tilde{\mathbf{\Sigma}} \in C^0(\mathscr{C}\times[0,S];\mathbf{M}_{2(n+1)}(\mathbb{R}))$ and $\tilde{\bm{h}} \in C^0([0,S];L^2(\mathring{\mathscr{C}};\mathbb{R}^{2(n+1)}))$ as in Eq. (46), and ICs $\bm{u}_0 \in L^2(\mathring{\mathscr{C}};\mathbb{R}^{2(n+1)})$, the PDE (47) admits a unique \emph{mild solution} $\bm{u} \in C^0([0,S];L^2(\mathring{\mathscr{C}};\mathbb{R}^{2(n+1)}))$. Additionally, if $\tilde{\mathbf{\Sigma}} \in C^1(\mathscr{C}\times[0,S];\mathbf{M}_{2(n+1)}(\mathbb{R}))$, $\tilde{\bm{h}} \in C^1([0,S];L^2(\mathring{\mathscr{C}};\mathbb{R}^{2(n+1)}))$, and the IC $\bm{u}_0 \in \mathscr{D}(\mathscr{A})$, with $\mathscr{D}(\mathscr{A}) \triangleq \{\bm{v} \in L^2(\mathring{\mathscr{C}};\mathbb{R}^{2(n+1)}) \mathrel{|} \pd{\bm{v}}{x} \in  L^2(\mathring{\mathscr{C}};\mathbb{R}^{2(n+1)}), \; \eval[0]{\bm{v}}_{\mathscr{L}} = \bm{0}\}$, the solution is \emph{classical}, that is, $\bm{u} \in C^1([0,S];\allowbreak L^2(\mathring{\mathscr{C}};\mathbb{R}^{2(n+1)})) \cap C^0([0,S];\mathscr{D}(\mathscr{A}))$.

\item The statement of Theorem 4.2 was modified as follows:

Suppose that $\bar{\bm{V}} \in C^1(\mathscr{C};\mathbb{R}^2)$ reads as in Eq. (49), and that $\mathring{\mathscr{C}}\subset \mathbb{R}^2$ is bounded, with boundary $\partial \mathscr{C}$ piecewise $C^1$. Then, for all $\tilde{\mathbf{\Sigma}}_{\varphi} \in C^0(\mathscr{C}\times[0,S];\mathbf{M}_{2(n+1)}(\mathbb{R}))$ and $\tilde{\bm{h}} \in C^0([0,S];L^2(\mathring{\mathscr{C}};\mathbb{R}^{2(n+1)}))$ as in Eq. (46), and ICs $\bm{u}_0 \in L^2(\mathring{\mathscr{C}};\mathbb{R}^{2(n+1)})$, the PDE (48) admits a unique mild solution $\bm{u} \in C^0([0,S];L^2(\mathring{\mathscr{C}};\mathbb{R}^{2(n+1)}))$. Additionally, if $\tilde{\mathbf{\Sigma}}_{\varphi} \in C^1(\mathscr{C}\times[0,S];\mathbf{M}_{2(n+1)}(\mathbb{R}))$, $\tilde{\bm{h}} \in C^1([0,S];L^2(\mathring{\mathscr{C}};\mathbb{R}^{2(n+1)}))$, and the IC $\bm{u}_0 \in \mathscr{D}(\mathscr{A})$, with $\mathscr{D}(\mathscr{A}) \triangleq \{\bm{v} \in L^2(\mathring{\mathscr{C}};\mathbb{R}^{2(n+1)}) \mathrel{|} (\bar{\bm{V}}\cdot\nabla_{\bm{x}})\bm{v} \in  L^2(\mathring{\mathscr{C}};\mathbb{R}^{2(n+1)}), \; \eval[0]{\bm{v}}_{\mathscr{L}} = \bm{0}\}$, the solution is classical, that is, $\bm{u} \in C^1([0,S];L^2(\mathring{\mathscr{C}};\mathbb{R}^{2(n+1)})) \cap C^0([0,S];\mathscr{D}(\mathscr{A}))$.

\item Throughout all the manuscript, the condition "$\bm{u}_0 \in H^1(\mathring{\mathscr{C}};\mathbb{R}^{2(n+1)})$ satisfying the BC" was replaced by "$\bm{u}_0 \in \mathscr{D}(\mathscr{A})$".
\end{enumerate}

\end{document}